\documentclass[twocolumn]{aastex63}
\usepackage{graphicx}	
\usepackage{amsmath}	
\usepackage{amssymb}	
\usepackage{physics}
\usepackage{enumitem}
\usepackage{bbm}
\usepackage{color}
\usepackage{natbib}
\usepackage{hyperref}

\newcommand{\LOIII}{$L_\mathrm{[O\,III]}$}
\newcommand{\LOII}{$L_\mathrm{[O\,II]}$}

\begin{document}

\title{Gas Content Regulates the Lifecycle of Star Formation and Black Hole Accretion in Galaxies}
\shortauthors{Yesuf \& Ho}
\author{Hassen M. Yesuf}
\affiliation{Kavli Institute for Astronomy and Astrophysics, Peking University, Beijing 100871, China}
\affiliation{Kavli Institute for the Physics and Mathematics of the Universe, The University of Tokyo, Kashiwa, Japan 277-8583}
\author{Luis C. Ho}
\affiliation{Kavli Institute for Astronomy and Astrophysics, Peking University, Beijing 100871, China}
\affiliation{Department of Astronomy, School of Physics, Peking University, 
Beijing 100871, China}

\begin{abstract}

Active galactic nucleus (AGN) feedback is expected to impact the amount of cold gas in galaxies by driving strong galactic winds, by preventing external gas inflows, or by changing the thermodynamical state of the gas. 
We use molecular gas mass estimates based on dust absorption (H$\alpha$/H$\beta$) to study gas content of large samples of type 2 AGN host galaxies in comparison with inactive galaxies. Using sparse principal component and clustering analysis, we analyze a suite of stellar and structural parameters of $\sim 27,100$ face-on, central galaxies at redshift $z = 0.02-0.15$ and with stellar mass $M_\star \approx 10^{10}-2\times 10^{11}\,M_\odot$. We identify four galaxy groups of similar mass and morphology (mass surface density, velocity dispersion, concentration, and S\'{e}rsic index) that can be evolutionarily linked through a lifecycle wherein gas content mediates their star formation rate (SFR) and level of AGN activity. Galaxies first consume their gas mostly through bursty star formation, then enter into a transition phase of intermediate gas richness in which star formation and AGNs coexist, before settling into retirement as gas-poor, quiescent systems with residual levels of AGN activity (LINERs). Strongly accreting black holes (Seyferts) live in gas-rich, star-forming hosts, but neither their gas reservoir nor their ability to form stars seem to be impacted \emph{instantaneously} (timescales $\lesssim 0.5$\,Gyr) by AGN feedback. Our results are inconsistent with AGN feedback models that predict that central, bulge-dominated, Seyfert-like AGNs in massive galaxies have significantly lower molecular gas fractions compared to inactive galaxies of similar mass, morphology, and SFR.
\end{abstract}


\section{INTRODUCTION}

There are two main types of galaxies in the local Universe: star-forming, gas-rich spiral galaxies, and quiescent, gas-poor elliptical and lenticular galaxies. Nearly all massive local galaxies harbor supermassive black holes ($\sim 10^6-10^{10}\,M_\odot$) at their centers. Through feeding and feedback mechanisms, they evolve with their host galaxies, and perhaps modulate each other's growth \citep{Kormendy+13}. 

The exact nature of (feedback) mechanisms that transform galaxies from star-forming to quiescent is unknown. Star formation can be quenched by removing or heating gas in galaxies or their surrounding halos \citep{Dekel+86,Silk+98,DiMatteo+05,Dekel+06,Hopkins+06, Martig+09,Peng+10,Fabian+12, Peng+15,Dubois+16,Pillepich+18}, either through external mechanisms that are governed by the dark matter halo or the environment, or through internally driven processes such as AGN feedback, stellar feedback, and morphological quenching.

Studying gas in the interstellar medium is critical for understanding the coevolution between supermassive black holes and their host galaxies. However, directly measuring gas content using radio telescopes for large, representative galaxy samples is difficult and time-consuming. Thus, to date, small and heterogenous samples of AGN hosts with cold gas measurements have yielded mixed results on the impact of AGN feedback. While some studies report suppressed gas content in AGNs \citep{Haan+08, Brusa+15, Kakkad+17, Perna+18}, others find that active galaxies have similar or even higher gas content relative to inactive galaxies \citep{Maiolino+97, Bertram+07,Ho+08, Fabello+11,Gereb+15,
Saintonge+17,Husemann+17,Yesuf+17,Rosario+18,Ellison+18,Shangguan+18,Shangguan+19,ZhuangHo20,Jarvis+20}.  

We recently proposed a cost-effective method to predict molecular gas masses using dust absorption as inferred from the Balmer decrement \citep{Yesuf+19}. Here we apply this method, in tandem with a multitude of other galaxy properties, to study the evolution of the molecular gas content of large samples of AGN host galaxies in comparison with inactive galaxies. Although it is important to test specific model predictions \citep[e.g.,][]{Harrison+17,Scholtz+18}, we take a data-driven approach and do not aim to test specific AGN feedback models. To our knowledge, comprehensive, quantitative predictions that can be compared with our approach and observations have not been made yet. It can be inferred, nevertheless, that different models make generally different predictions of the relationships between AGN activity, gas fractions, star formation rates (SFRs), and galaxy morphology. We hope our approach and general results inspire follow-up, direct comparisons.

Accretion onto a black hole may release $\sim 10$\% of its rest mass energy, which is comparable to the binding energy of gas in the halo \citep[e.g.,][]{Silk+98, Fabian+12}. Even if a small fraction of the energy can be coupled to the gas, an AGN can have very significant effects on the evolution of its host galaxy \citep{Fabian+12, Kormendy+13}. Depending on the black hole accretion rate, AGN feedback is modeled using two main modes \citep[although see][]{Schaye+15}. At high accretion rates, a radiative mode (also called quasar mode) is used \citep[e.g.,][]{DiMatteo+05,Hopkins+06}, while at low accretion rates, a kinetic mode (also called radio mode), producing high-velocity winds and/or radio jets, is used \citep[e.g.,][]{Croton+06,Weinberger+17}. Galaxy formation models that do not include AGN feedback form stars too efficiently and fail to reproduce even basic properties of massive galaxies.

Previous work related to IllustrisTNG simulations suggest that the transition between star-forming and quenched galaxies is tied to the onset of kinetic mode feedback \citep{Weinberger+17, Habouzit+19, Nelson+19,Terrazas+20, Zinger+20}. Kinetic feedback removes gas from the star-forming regions as well as leads to hotter, more dilute and higher entropy circumgalactic medium (CGM) with long cooling times \citep{Zinger+20}. Hence, the AGN feedback in TNG simulations ejects and heats up the gas within and around galaxies. The cumulative wind energy from a low-accretion rate black hole is crucial to produce gas-deficient, compact, quiescent galaxies, above a threshold of $M_\bullet \approx 2 \times 10^8\,M_\odot$ or $M_\star \approx 3 \times 10^{10}\,M_\odot$. Below this threshold most simulated central galaxies are star-forming, and above the threshold most are quiescent \citep{Terrazas+20,Zinger+20}.

In contrast, in the EAGLE simulations \citep{Bower+17} it is proposed that star formation-driven outflows regulate the amount of gas reaching the black holes and set the characteristic stellar mass $M_\star \approx 3 \times 10^{10}\,M_\odot$ or halo mass $M_{h} \sim 10^{12}\,M_\odot$. Outflows are efficient in low-mass galaxies, making it difficult for the build-up of cool gas in the central regions of galaxies. In massive galaxies, which are surrounded by hot halos, outflows cease to be buoyant and are less efficient, and thus the unrestrained central gas build up leads to rapid black hole growth. In turn, the black holes become effective at heating the halo gas, disrupt the infalling cool gas supply, and slowly starve the galaxies of the fuel for future star formation.

Compared to those in the EAGLE simulations, the winds in TNG simulations eject significantly larger amounts of gas from the centers of massive galaxies \citep{Nelson+19,Mitchell+20}. Likewise, EAGLE simulations do not predict a negative trend between SSFR and AGN luminosity; AGN feedback does not reduce galaxy-wide instantaneous SFRs of luminous, simulated AGN host galaxies \citep{Scholtz+18}. The signature of AGN feedback is instead imprinted on the overall SSFR distributions of massive galaxies. Thus, there may be implicit/subtle relationships between the SFRs, gas fractions and AGN luminosities since the timescale of an AGN episode is shorter than the timescale for the suppression of star formation by possibly multiple AGN episodes \citep{Harrison+17,McAlpine+17,Scholtz+18,Schulze+19}.

In Section~\ref{sec:data}, we describe the data, sample selection and the methods used to identify our proposed evolutionary sequence. In Section~\ref{sec:res}, we present relationships among nuclear activity, gas, stellar, and structural properties, along the sequence. Section~\ref{sec:disc} presents a discussion of our results. A summary of this work and its main conclusions are given in Section~\ref{sec:conc}.

\section{DATA AND  METHODOLOGY}\label{sec:data}

\subsection{Data}

Our galaxy sample is taken from the Seventh Data Release of the Sloan Digital Sky Survey \citep[SDSS DR7;][]{Abazajian+09,Alam+15}. The publicly available Catalog Archive Server (CAS) \footnote{http://skyserver.sdss.org/casjobs/ \\ We use CAS in the context of data release 13 (DR13) to retrieve various measurements in different catalogs. However, the sample of galaxies we use is restricted to those in DR7, because the axis ratio and S\'{e}rsic index measurements from \citet{Simard+11} are only available for DR7 galaxies. The following tables are queried: {\tt photoobjall, galSpecIndx, galSpecInfo, galSpecLine, galSpecExtra, and {\tt specDR7.}}} is used to collate some of the measurements used in this work (e.g., emission-line fluxes and spectral indices). These data are supplemented with stellar mass and SFR from version 2 of the GALEX-SDSS-WISE Legacy Catalog\footnote{http://pages.iu.edu/$\sim$salims/gswlc/} \citep[GSWLC-2;][]{Salim+16,Salim+18}, along with structural parameters (S\'{e}rsic index and ellipticity/axial ratio) derived from single-component S\'{e}rsic function fits \citep{Simard+11}, and environmental information from the SDSS group catalog \citep{Lim+17}. To assess the importance of bars on AGN fueling, we use a bar classification catalog based on machine learning \citep{Dominguez+18}. The deep learning model was trained on visual classifications of SDSS images by expert astronomers \citep{Nair+10}. For the $i$th galaxy in a given sample of size $n$, the machine classification outputs a probability $p_i$ that it has a bar. Assuming a Poisson-binomial distribution for the number of barred galaxies in the given sample, we estimate its mean bar fraction, $\mu_{f_\mathrm{bar}} = \frac{\sum^{n_i}_{i=1} p_i}{n}$, and its standard deviation, $\sigma_{f_\mathrm{bar}} =\frac{\sqrt{\sum^{n_i}_{i=1} p_i(1-p_i)}}{n}$. The binomial distribution is a special case of the Poisson-binomial distribution, when all probabilities are the same.

\begin{figure*}
\includegraphics[width=0.49\linewidth]{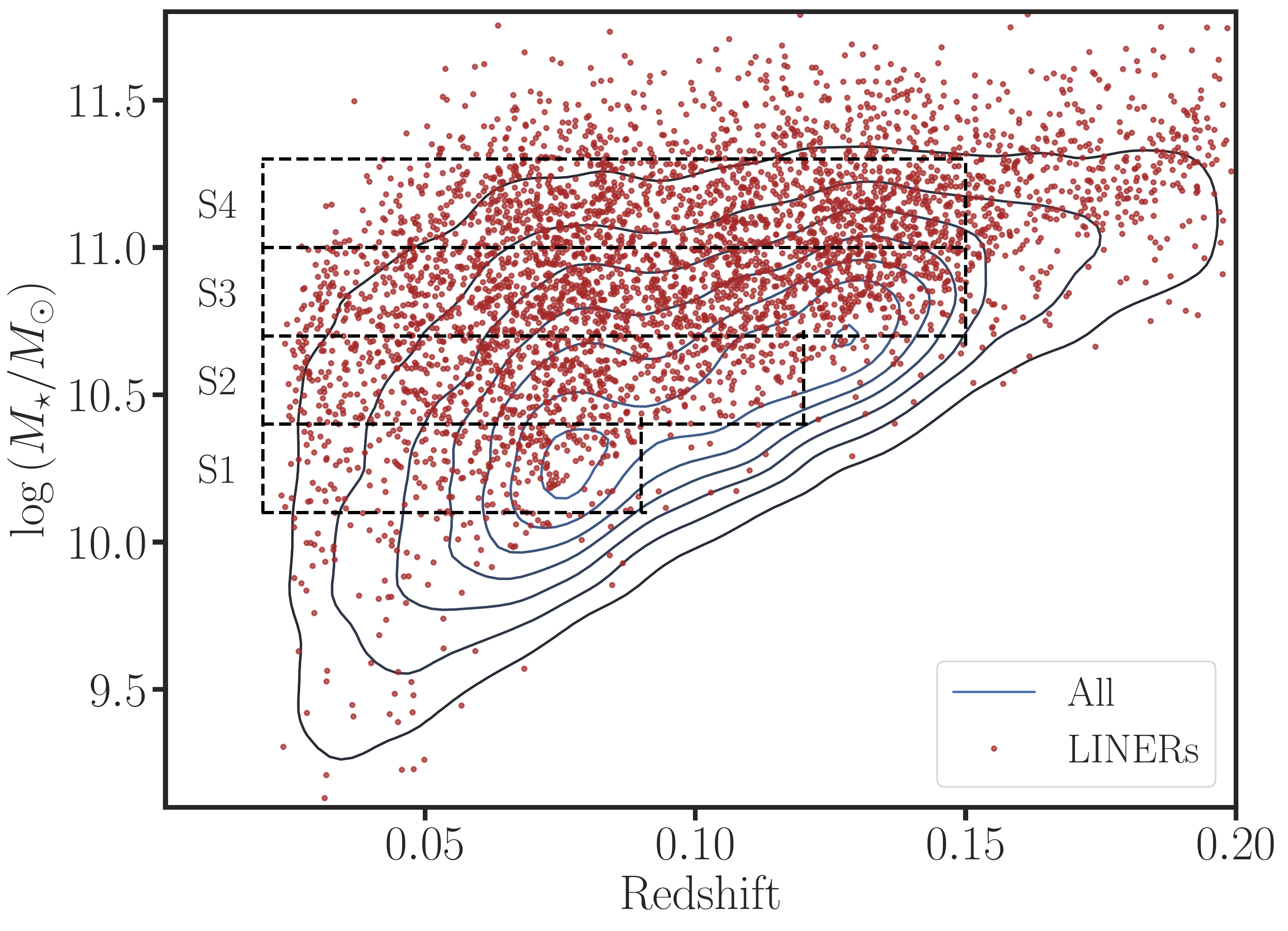}
\includegraphics[width=0.48\linewidth]{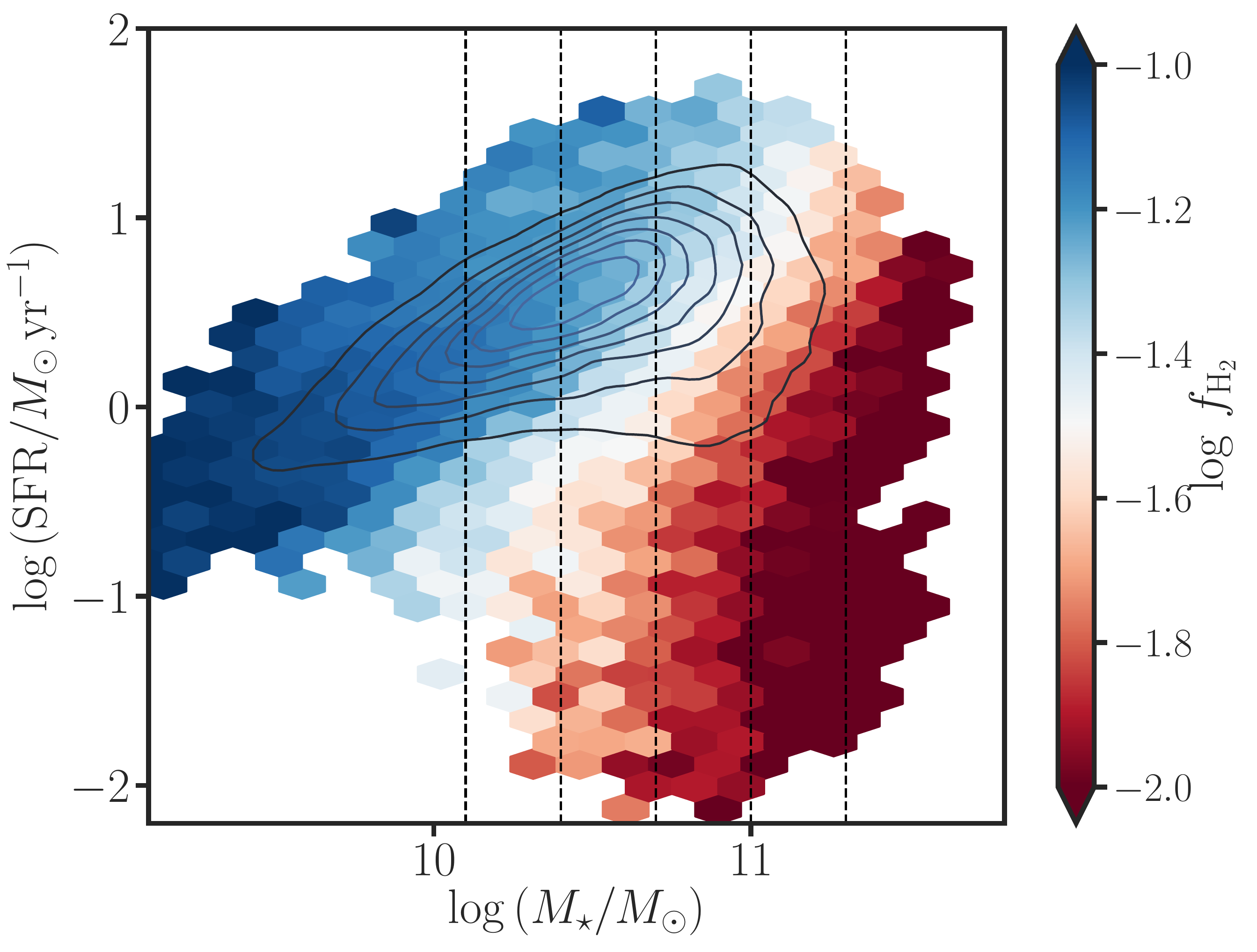}
\caption{Four approximately ``volume-limited" samples (S1-S4, left panel), for which we follow the evolution of SFR, molecular  gas fraction ($f_\mathrm{H_2}$), and AGN activity in galaxies of similar stellar mass ($M_\star$) and morphology. The mass ranges span galaxies transitioning from gas-rich/star-forming to gas-poor/quiescent galaxies (right panel, the color-coding is by the median molecular gas mass fractions, which are estimated from H$\alpha$/H$\beta$ following \citet{Yesuf+19}.
 \label{fig:mz}}
\end{figure*}

We define the effective mass surface density as $\Sigma_\star = M_\star/(2\pi R_{50,z}^2)$, where $M_\star$ is the total stellar mass, and $R_{50,z}$ is  the $z$-band half-light radius. The concentration index is defined as the ratio of 90\% Petrosian radius to the 50\% Petrosian radius in the $r$ band, $C = R_{90}/R_{50}$. 

In our adopted group catalog \footnote{https://gax.sjtu.edu.cn/data/Group.html \\ The group catalog is based on SDSS DR13.} \citep{Lim+17}, the groups were identified with an iterative halo-based group finder \citep{Yang+07}, which used the stellar mass of a central galaxy as well as the stellar mass difference between the central galaxy and the $n$th most massive satellite as halo-mass proxies. An abundance matching technique was used to assign final halo masses to distinct groups. For groups that were not assigned masses by the abundance matching, due to mass incompleteness, halo masses were assigned based on the mean relation between the halo mass and its proxies obtained from the group finder. Mock galaxy samples constructed from EAGLE simulations were used to test and calibrate the group finder.

\begin{figure*}
\centering
\includegraphics[width=0.95\linewidth]{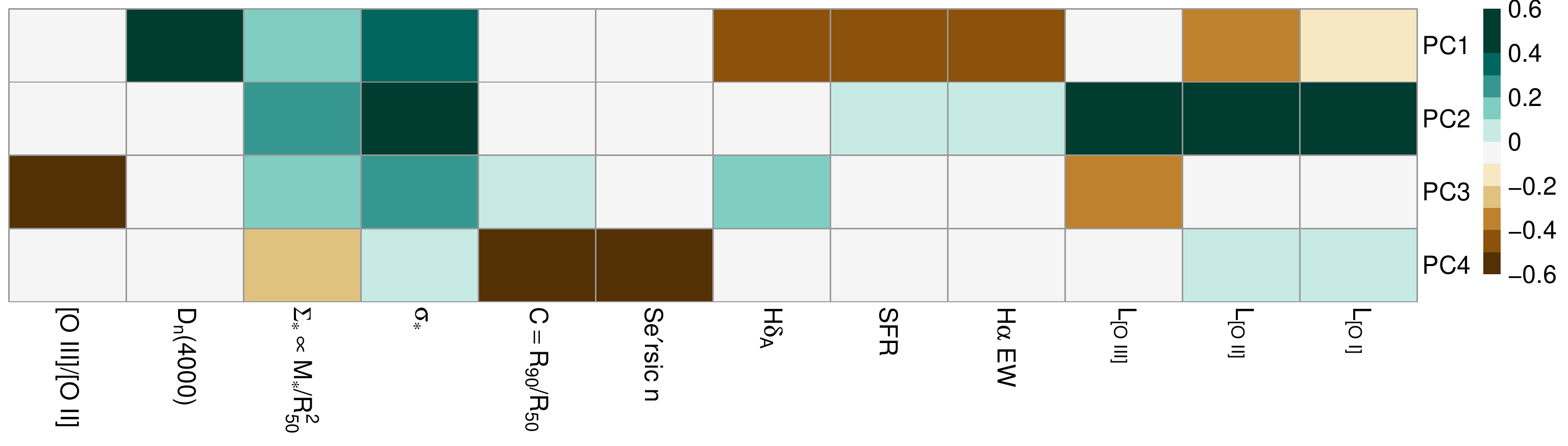}
\caption{The 12 parameters used in the SPCA analysis. The colorbar indicates the weights given to parameters in constructing the PCs for the S2 sample using $\alpha=\beta = 10^{-3}$. PC1 is mainly sensitive to variation of SFR with stellar age [$D_n(4000)$], PC2 to black hole accretion in bulges, PC3 to the variation of ionization parameter (contrasts [\ion{O}{3}], [\ion{O}{2}], and morphology), and PC4 to the variation of morphology (contrasts $C$ and $n$ with stellar velocity dispersion, $\sigma_\star$). The four PCs account for about 85\% of the variance. \label{fig:pc}}
\end{figure*}

\begin{figure*}
\includegraphics[scale=0.3]{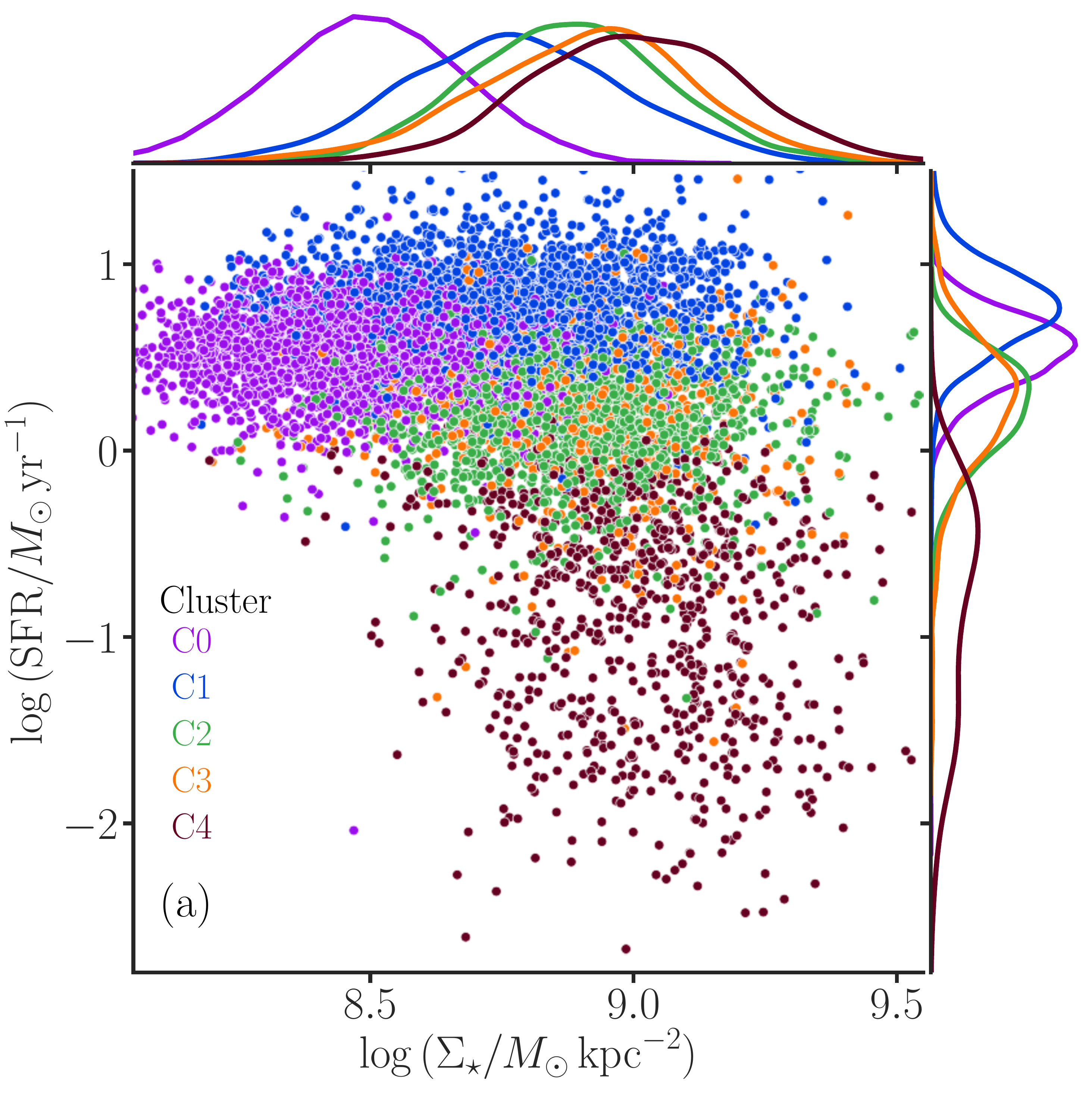}
\includegraphics[scale=0.3]{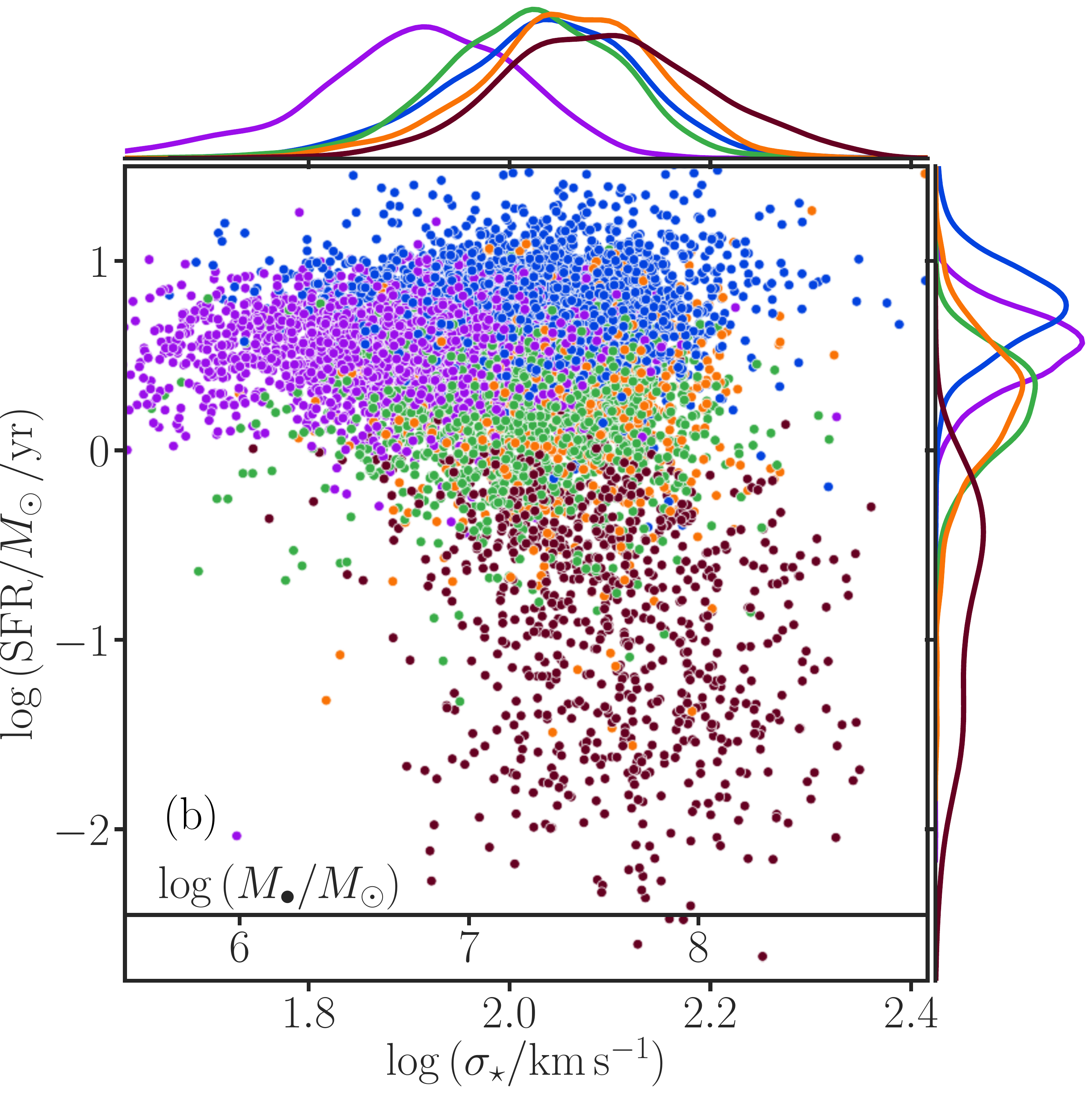}
\includegraphics[scale=0.3]{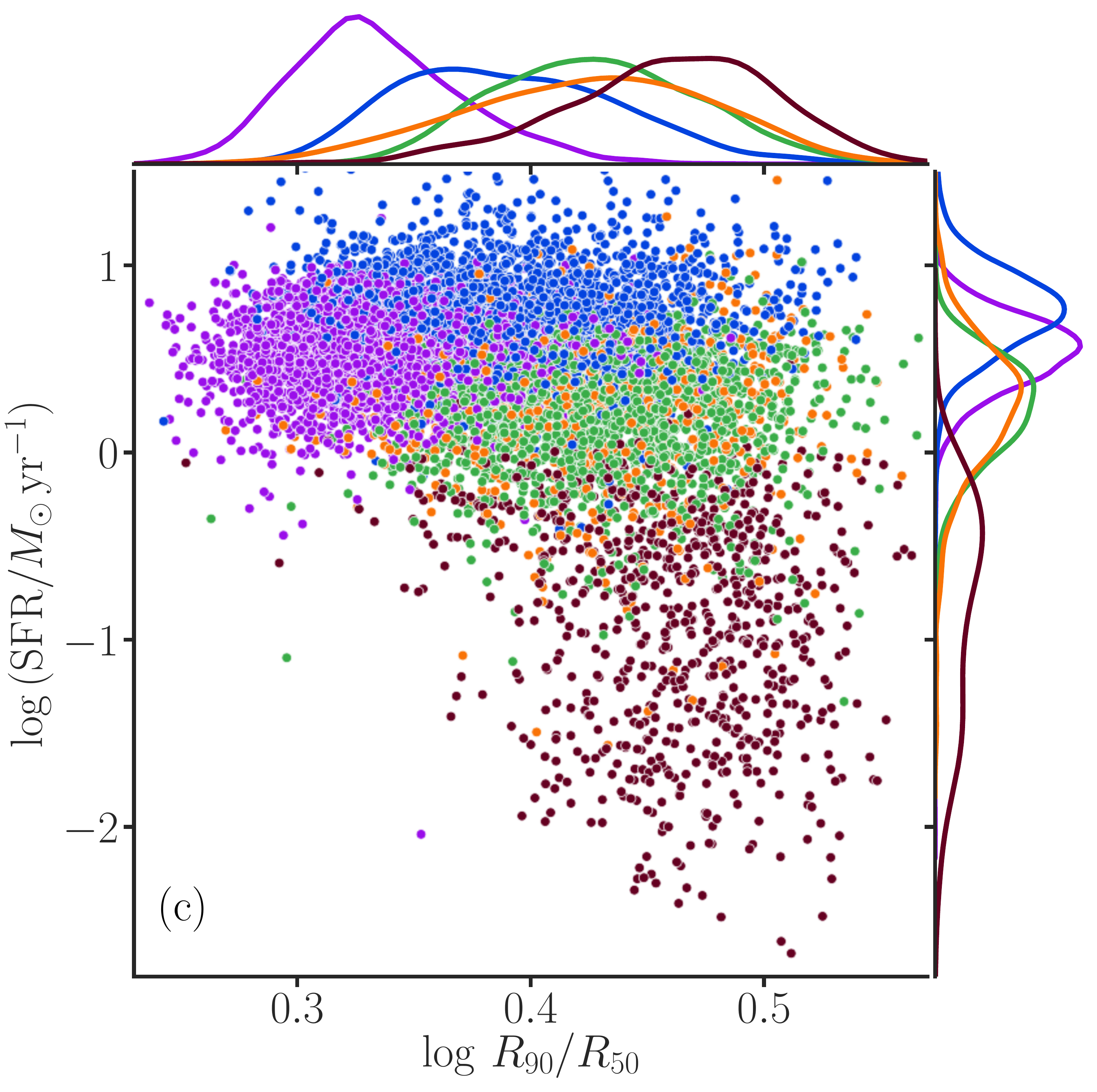}
\hfill
\includegraphics[scale=0.3]{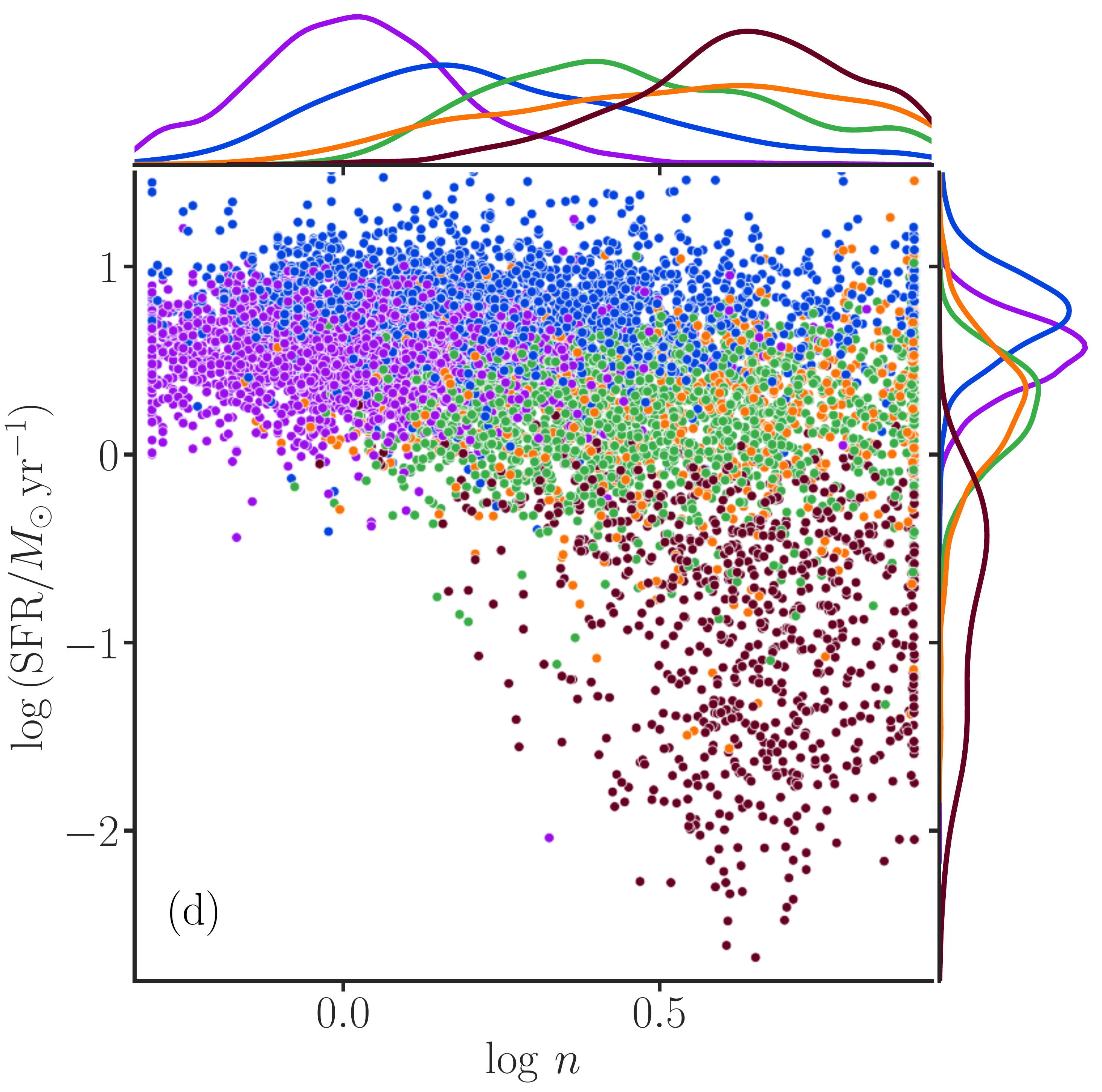}
\caption{Variation of SFR versus (a) the stellar mass surface density ($\Sigma_\star$), (b) stellar velocity dispersion (and the black hole mass estimates using the \citet{Kormendy+13} relation of classical bulges and ellipticals), (c) the concentration index, $C=R_{90}/R_{50}$, where $R_{90}$ and $R_{50}$ are the radii that enclose 90\% and 50\% of the light, and (d) S\'{e}rsic index for galaxies with stellar mass $\log\,(M_\star/M_\odot) = 10.4 -10.7$. The colored points are the five clusters from our SPCA/clustering analysis. The C0 (violet) cluster is structurally very different from the rest, and it may not be their immediate progenitor. We will show that the other clusters evolve from C1 (blue) to C4 (maroon). The margin curves show the kernel density estimates of the quantities plotted. Unlike $\sigma_\star$, the light-based $C$ and $n$ may significantly change as the disk stellar populations fade.\label{fig:morph}}
\end{figure*}

\subsection{The main sample selection}

The following steps are taken to select the main sample used in this work:

\begin{itemize}
\item Select galaxies in the redshift range $z = 0.02-0.15$ and stellar mass range $\log\,(M_\star/M_\odot) = 10.1 -11.3$ from the four narrow-mass samples shown in Figure~\ref{fig:mz}. The four samples, denoted by S1 to S4, have stellar mass $\log \, (M_\star/M_\odot) = 10.1-10.4$, $10.4-10.7$, $10.7-11.0$, and $11.0-11.3$, respectively, and $z = 0.02-0.09$, $0.02-0.12$, $0.02-0.15$, and $0.02-0.15$, respectively.
\item Include only central galaxies (not satellites) in the group catalog \citep{Lim+17}. The results are similar if we only use isolated galaxies, with no nearby neighbors.
\item Exclude edge-on galaxies (axis ratio $b/a < 0.5$). The orientation effect may complicate the estimation of gas mass from dust absorption in edge-on galaxies.
\item Exclude type 1 AGNs by restricting the velocity dispersions of the Balmer emission lines to values less than $\sigma_\mathrm{H\alpha} = 400$ km s$^{-1}$, below which very few type 1 AGNs exist \citep{GreeneHo2007}. For these objects, the AGN may significantly affect the measurements of the host galaxy properties.
\item Select galaxies with signal-to-noise ratio SNR $> 3$ for the flux densities of H$\alpha$, H$\beta$, [\ion{O}{3}]~$\lambda$5007, [\ion{O}{2}]~$\lambda\lambda$3726, 3729, and [\ion{O}{1}]\,$\lambda$6300 \footnote{If the requirement of the faint [\ion{O}{1}]\,$\lambda$6300 line is omitted, the sample size increases significantly but the results that do not depend on [\ion{O}{1}] do not change significantly.}. These criteria ensure that the molecular gas mass, AGN luminosity, and the ionization parameter are reliable.
\item Select galaxies with well-measured (fiber) stellar velocity dispersion (SNR $> 3$).

\end{itemize}

The main results are similar for all four samples and we take the sample S2 [$\log\,(M_\star/M_\odot) = 10.4-10.7$] as the fiducial sample. To avoid redundancy, we selectively present some of the results for the other samples.

\subsection{Estimation of molecular gas masses}\label{ssec:gas}

We calculate the $V$-band dust absorption using the observed H$\alpha$/H$\beta$ ratio and the dust attenuation 
curve \citet[][see also Wild et al. 2011]{Charlot+00}, as follows:

\begin{equation}
Q_\lambda = 0.6\,(\lambda/5500)^{-1.3} + 0.4\,(\lambda/5500)^{-0.7}.
\end{equation}

\noindent
Assuming that the intrinsic Balmer decrement H$\alpha$/H$\beta$ = 2.86 for inactive galaxies and H$\alpha$/H$\beta$ = 3.1 for AGNs \citep[e.g.,][]{Ferland+83,Gaskell+84},

\begin{equation}\label{eq:AV}
 A_{V} = \frac{2.5}{(Q_{4861} - Q_{6563})} \times \log \frac{\mathrm{H}\alpha/\mathrm{H}\beta}{3.1 \,\mathrm{or} \, 2.86},
\end{equation}

\noindent where $Q_{4861} - Q_{6563} = 0.31$. If the observed ratio of an object is below the intrinsic ratio (2.86 or 3.1), we set $A_V = 0$ mag\footnote{This might cause the object's predicted gas mass to be lower by up to $\sim 0.1$ dex than it actually is. Such small changes are not important for our application. An object with H$\alpha$/H$\beta < 2.9-3.1$ might have $\log\, (M_\mathrm{H_2}/M_\odot) \lesssim 8.5-8.8$.}. For reference, $A_V = [0.5,1,1.5, 2, 3]$ mag corresponds to $\mathrm{H\alpha/H\beta} \approx [3.3, 3.8, 4.4, 5.1, 6.8]$.

We estimate the molecular gas mass using our empirical (median) estimator presented in our previous work \citep{Yesuf+19}. In particular, we use nebular dust absorption, the average gas-phase metallicity, which is inferred from the stellar mass-metallicity relation \citep{Tremonti+04}, and the half-light radius to estimate $M_\mathrm{H_2}$, with uncertainty of $\sim 0.4$ dex. The results on the gas masses are similar if we use the dust absorption derived from fitting the stellar continuum alone or in combination with H$\alpha$/H$\beta$.

\subsection{Estimation of black hole accretion rate}\label{ssec:gas}

We combine dust-corrected [\ion{O}{3}]\,$\lambda$\,5007 and  [\ion{O}{1}]\,$\lambda$\,6003 luminosities to estimate the total bolometric luminosity ($L_\mathrm{bol}$) of the AGN, with an uncertainty of $\sim 0.5$ dex \citep{Netzer09}.  We calculate the Eddington luminosity, $L_{\rm Edd}\,=\,1.26 \times 10^{38} \left(M_\bullet/M_{\odot}\right)\,{\rm erg~s^{-1}}$, using black hole masses ($M_\bullet$) derived from the fiber stellar velocity dispersions \citep{Kormendy+13}. The SDSS fiber spans different regions of galaxies for different redshifts; we do not correct for this effect in our black hole estimates, which has a nominal uncertainty of $\sim 0.3$ dex. Thus, the resulting Eddington ratio ($\lambda_\mathrm{Edd} \equiv L_\mathrm{bol}/L_\mathrm{Edd}$), a rough approximation of the mass accretion rate, has an uncertainty of $\sim 0.6$ dex.

\subsection{SPCA \& clustering analysis in narrow mass ranges}

Star formation and AGN activities evolve on timescales that are much shorter than a galaxy merger timescale \citep{Fang+13,Yesuf+14,Rodriguez-Gomez+15,Trayford+16, Hahn+17}. Most local star-forming galaxies are unlikely to grow more than 0.3 dex in stellar (or black hole) mass by mergers in the time it takes them to cease forming stars. Galaxies that have similar masses but different internal mass distributions and environments are also unlikely to be immediate progenitors and descendants. Thus, to select candidate galaxies that are likely to be directly evolutionarily linked, we choose central, massive galaxies in four narrow stellar mass ranges. 

Even in a restricted mass range and environment, galaxies are diverse in their gas content, SFR, and star formation history (SFH), stellar population age, internal kinematics, morphology (mass/light distribution), and AGN activity. Some of these properties not only correlate with each other, but also may be sensitive to multiple, slightly different indicators. To explore statistically the evolution of central, massive galaxies in the four samples of approximately constant mass, we perform sparse principle component analysis (SPCA) and clustering analysis of 12 variables that describe the morphology and the AGN and star formation activities of galaxies (Figure~\ref{fig:pc}). PCA seeks a reduced set of new, uncorrelated variables, called principal components (PCs), which are linear combinations of the original variables. PCs sequentially capture the maximum variance in the data, and the first PC accounts for the highest variance. Because each PC is a linear combination of all variables, usually with non-zero weights, it is generally difficult to interpret the derived PCs. SPCA is a modern variant of PCA in which the PCs are derived from only a few of the most important variables. Furthermore, the clustering analysis aims to partition the reduced set of variables into ``clusters,'' such that galaxies assigned to the same cluster are less dissimilar than those in different clusters\footnote{The ``clusters'' identified through the clustering analysis should not be confused with gravitationally bound galaxy clusters.}. After efficiently identifying the clusters using all the relevant information, we visualize how they distribute in two-dimensional spaces of well-known diagrams to reveal evolutionary paths. Stellar spectral indices are used to gauge the average stellar population age of each cluster, and to determine its position on the evolutionary sequence.

We implement SPCA using the \texttt{sparsepca} package \citep{Erichson+18} in \texttt{R}. SPCA can be formulated as a regularized regression-type problem \citep{Zou+06}. Given the data matrix $\mathbf{X}$, SPCA attempts to minimize the function $\frac{1}{2} \lVert \mathbf{X} -\mathbf{XBA^{\intercal}} \rVert^2 + \alpha \lVert \mathbf{B} \rVert +\frac{1}{2}\beta \lVert \mathbf{B} \rVert^2$, subject to $\mathbf{A^\intercal A}= \mathbf{I}$, where $\mathbf{B}$ is a sparse weight matrix, $\mathbf{A}$ is an orthonormal matrix, and $\mathbf{I}$ is the identity matrix. So, the PC matrix is $\mathbf{XB}$. The last two terms of the function are known as the elastic net regularizer. The first term shrinks the weights toward zero if $\alpha$ is large enough. We use four PC components, accounting for $\sim 85$\% of the variance, and $\alpha = \beta = 10^{-3}$. The main results do not change significantly if we instead use a value of $10^{-4}$ for both $\alpha$ and $\beta$. We also center and scale the variables to have unit variance.

We perform $k$-means clustering of the PC scores in \texttt{R} to group the data into $k=5$ or $k=6$ clusters. One of the simplest and most popular clustering algorithms, $k$-means aims to iteratively partition a given dataset of $n$ observations into $k$ optimal clusters. In the end, each observation is assigned to the nearest cluster, such that the squared Euclidian distances of the observations from the cluster centers are minimized. Determining objectively the optimal number of clusters in a dataset is not an easy problem. Different metrics (the gap statistics, the elbow method, etc.) indicate $k \approx 3-6$. Using $k=3$ gives similar results in which three of the four bulge-dominated groups (C1-C3) are merged together. While this reinforces the fact that these clusters are evolutionary related, this choice, unlike $k=5$ or $k=6$, does not provide sufficiently fine distinction to illustrate the evolutionary stages discussed in Section~3.

\subsection{Toy stellar population models} To illustrate how a recent burst of star formation evolves in the $D_n(4000)-\mathrm{H}\delta_A$ diagram \citep{Balogh+99,Kauffmann+03}, we use the updated version of the \citet{Bruzual+03} stellar population code to model the SFHs of galaxies as a superposition of an old stellar population initially formed at time $t= 5$ Gyr ($z \approx 1.2$) and a young population that formed in a recent burst at $t = 12$ Gyr ($z \approx 0.12$). The recent burst fraction is 5\%, 10\% or 30\%. The old population formed following a delayed exponential SFH of the form $\psi \propto t\exp(-t/\tau_1)$ with e-folding time $\tau_1 = 1$ Gyr, while the young population has SFH of the form $\psi \propto \exp(-t/\tau_2)$ with $\tau_2= 0.1$ Gyr \citep{Yesuf+14}. The single stellar population models assume a \citet{Chabrier+03} stellar initial mass function and solar metallicity before the recent burst and 2.5 times solar after the burst.

\section{RESULTS}\label{sec:res}

\subsection{Relationships among nuclear activity, gas, stellar, and structural properties}

Figure~\ref{fig:morph} shows how the morphology (mass surface density, velocity dispersion, concentration or S\'{e}rsic $n$) of the five clusters relates to their SFRs. The cluster C0 mostly contains less dense ($\Sigma_\star \lesssim 10^{8.5}\,M_\odot\,\mathrm{kpc}^{-2}$), lower velocity dispersion ($\sigma_\star < 100\,\mathrm{km\,s}^{-2}$), and less concentrated ($C < 2.5$ and $n < 2$) galaxies compared to the other clusters. The galaxies in C0 are exclusively star-forming galaxies. The clusters from C1 to C4 span a wide range in SFR and gas mass, at a constant morphology. C1 has the highest average SFR while C4 has the lowest average SFR; the SFRs of C2 and C3 are intermediate between those of C1 and C4. The galaxies in C0 are likely pseudo bulges, and the rest classical bulges or ellipticals \citep{Yesuf+20a}. Although our current theoretical understanding of bulges is incomplete, the common view is that classical bulges and ellipticals are formed mainly by early dissipative mergers, while pseudo bulges are formed by disk-related, internal secular processes \citep{Kormendy2004,Brooks+16}. Black holes also correlate differently with the properties of the two bulge types \citep{Kormendy+13}. The correlations with pseudo bulge properties are weak and imply no close coevolution. Thus, galaxies in C0 may have undergone different formation histories than those in the other clusters. While they may eventually evolve to become galaxies in C1, this is unlikely to happen in a short period of time.

Simulations show that the evolution tracks in the $\mathrm{SFR}-\Sigma_\star$ plane (Figure~\ref{fig:morph}) have a characteristic ``L'' shape \citep{Zolotov+15,Tacchella+16, Choi+18}. Gas consumption by star formation and gas ejection by stellar and supernova feedback lead to quenching of star formation at a critical value of $\Sigma_\star$. Further gas accretion is prevented because the galaxies have reached the critical halo mass to form a stable viral shock, which keeps the CGM at the virial temperature \citep{Dekel+06}. The halo mass for our fiducial sample is $\log \,(M_{h}/h^{-1}M_\odot) = 12 \pm 0.1$ \citep{Lim+17}.  In the EAGLE simulations, \citet{Correa+19} found that the time when galaxies move to the red sequence depends on their morphology, and AGN feedback is important for quenching central ellipticals, but not for disks. Next, we show the nature of the nuclear activities in groups C0 to C4.

\begin{figure*}
\centering
\includegraphics[scale=0.37]{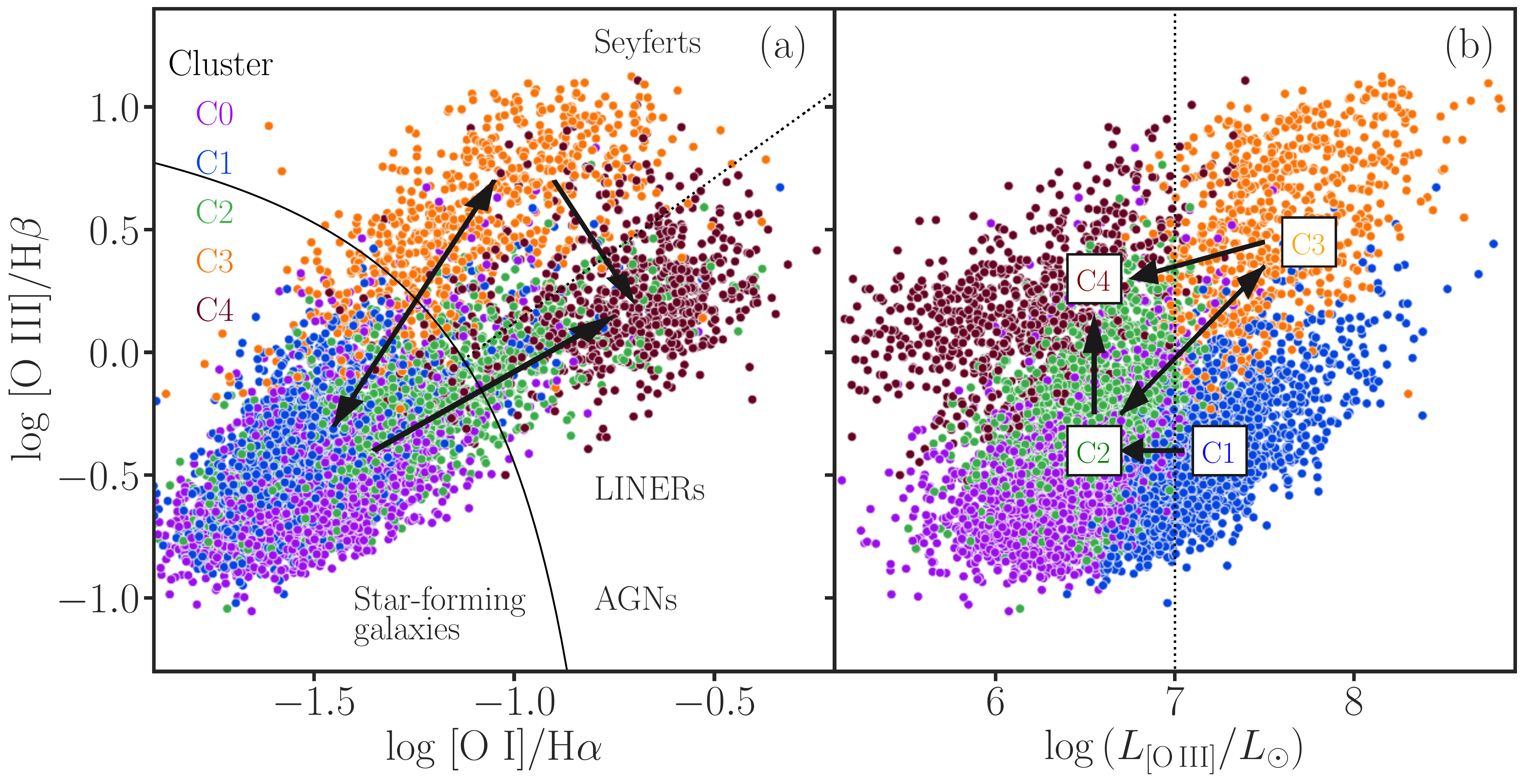}
\caption{(a) Evolution on spectral diagnostic diagrams for galaxies with stellar mass $\log\,(M_\star/M_\odot) = 10.4-10.7$. The solid curve demarcates the theoretically maximum boundary for star-forming galaxies; galaxies above this curve are dominated by AGN emission \citep{Kewley+01}. The dotted line separates strong AGNs (Seyferts) from weak AGNs (LINERs) \citep{Kauffmann+03,Kewley+06}. Without using the standard emission-line ratios, our clustering analysis identifies four clusters that broadly correspond to star-forming galaxies, Seyferts, and LINERs. The clusters evolve from C1 (blue) to C4 (maroon). In panel (b), C3 is distinguishable from C2 and C4 because it has stronger {\LOIII} from black hole accretion; likewise, C1 has higher SFRs and therefore produces stronger {\LOIII} than C2. \label{fig:bpt_evol}}
\end{figure*}

The optical emission-line intensity ratio diagnostic diagrams effectively discriminate whether the source of ionization in a galaxy arises from recently formed hot, massive stars or accretion onto a black hole \citep{Baldwin+81,Veilleux+87,Ho+97,Kewley+01,Kauffmann+03,Kewley+06}.  Figure~\ref{fig:bpt_evol}a shows the most sensitive of the line-ratio diagrams, involving [\ion{O}{3}]~$\lambda 5007$/H$\beta$ and [\ion{O}{1}]~$\lambda 6300$/H$\alpha$. AGN-dominated galaxies occupy the upper-right corner of the diagram, while star formation-dominated galaxies lie in the lower-left corner. Galaxies hosting AGNs lie on two branches associated with Seyferts and LINERs, whose difference in ionization stems from their difference in accretion rate \citep{Ho2009}. Our clusters occupy distinct loci normally assigned to standard galaxy spectral classes, even though the line-ratio classification criteria were {\it not}\ among the parameters used in the SPCA analysis. 

For the fiducial sample with stellar mass $\log\,(M_\star/M_\odot) = 10.4-10.7$, nearly all of the members in C1 (98\%) and 86\% of C2 are classified as star-forming galaxies, with the former systematically more luminous than the latter owing to their larger SFRs. By contrast, AGNs comprise most ($75\%-90\%$) of the galaxies in C3 and C4, predominantly Seyferts in C3 (74\%) and LINERs in C4 (74\%), whose difference in AGN power is reflected in their [O\,III] luminosity \citep{Heckman+04}. The four classes segregate quite cleanly in a plot of {\LOIII} versus [\ion{O}{3}]/H$\beta$ (Figure~\ref{fig:bpt_evol}b). An additional, similar result for [\ion{N}{2}]~$\lambda 6584$/H$\alpha$ and [\ion{S}{2}]~$\lambda\lambda 6716, 6731$/H$\alpha$ diagrams is given in the Appendix.

\begin{deluxetable*}{cccccc}
\tabletypesize{\scriptsize}
\tabletypesize{\footnotesize}
\tablecolumns{6} 
\tablewidth{0pt}
\tablecaption{The Fraction of AGNs and Star-forming Galaxies \label{tbl:agn_frac}}
\tablehead{
\colhead{Sample} & \colhead{Cluster} & \colhead{Seyferts}  & \colhead{LINERs} & \colhead{Star-forming Galaxies} & \colhead{$\log \, (L_\mathrm{O\,III}/L_\odot)  > 7$}
}
\startdata
 & C0a & $0.003 \pm 0.001$ & $0.009 \pm 0.002$  & $0.988 \pm 0.003$ & $0.0$ \\ 
 & C0b & $0.0$ & $0.0$  & $1.0$ & $0.031 \pm 0.004$\\ 
 S1 & C1 & $0.019 \pm 0.004 $ & $0.012 \pm 0.004$  & $0.969 \pm 0.006$ & $0.482 \pm 0.016$ \\ 
 & C2 & $0.016 \pm 0.004 $ & $0.049 \pm 0.006$  & $0.935 \pm 0.007$ & $0.010 \pm 0.003$ \\ 
 & C3 & $0.701 \pm 0.026$& $0.020 \pm 0.008$  & $0.279 \pm 0.026$ &$0.631 \pm 0.028$ \\ 
 & C4 & $0.166 \pm 0.019$& $0.545 \pm 0.025$  & $0.288 \pm 0.023$ &$0.008 \pm 0.005$ \\ 
 \hline
 & C0 & $0.012 \pm 0.002$& $0.009 \pm 0.002$ & $0.979 \pm 0.003$ & $0.020 \pm 0.003$ \\ 
 & C1 & $0.010 \pm 0.002$ & $0.014 \pm 0.002$ & $0.976 \pm 0.003$ & $0.573 \pm 0.01$ \\ 
S2 & C2 & $0.035 \pm 0.004 $ & $0.108 \pm 0.007$  & $0.857 \pm 0.008$ & $0.040 \pm 0.005$ \\ 
 & C3 & $0.742 \pm 0.017 $ & $0.016 \pm 0.005$  & $0.242 \pm 0.016$ & $0.963 \pm 0.007$ \\ 
 & C4 & $0.168 \pm 0.013$& $0.737 \pm 0.016$  & $0.094 \pm 0.010$ &$0.029 \pm 0.006$ \\ 
 \hline
  & C0 & $0.036 \pm 0.004$ & $0.031\pm 0.003$  & $0.932 \pm 0.005$ &$0.239 \pm 0.009$ \\ 
 & C1 & $0.045 \pm 0.005$ & $0.036 \pm 0.004$  & $0.919 \pm 0.006$ & $0.930 \pm 0.006$\\ 
 S3 & C2 & $0.056 \pm 0.005 $ & $0.195 \pm 0.009$  & $0.749 \pm 0.010$ & $0.280 \pm 0.010$ \\ 
  & C3 & $0.802 \pm 0.012$ & $0.011 \pm 0.003$  & $0.187 \pm 0.012$ & $0.994 \pm 0.002$ \\ 
 & C4 & $0.128 \pm 0.009$ & $0.833 \pm 0.010$  & $0.039 \pm 0.005$ &$0.081 \pm 0.007$ \\ 
\hline
 & C0 & $0.107 \pm 0.011$ & $0.173 \pm 0.013$  & $0.720 \pm 0.015$ &$0.362 \pm 0.016$ \\ 
 & C1 & $0.079 \pm 0.010$ & $0.095 \pm 0.011$  & $0.826 \pm 0.014$ & $0.971 \pm 0.006$\\  
 S4 & C3 & $0.889 \pm 0.014$ & $0.021 \pm 0.06$  & $0.090 \pm 0.013$ & $0.996 \pm 0.003$ \\ 
 & C4a/C2 & $0.136 \pm 0.012 $ & $0.533 \pm 0.018$  & $0.331 \pm 0.017$ & $0.539 \pm 0.018$ \\
 & C4b & $0.069 \pm 0.009$ & $0.917 \pm 0.009$  & $0.014 \pm 0.004$ &$0.053 \pm 0.008$ \\ 
\enddata
\tablecomments{There some variations between clusters in different mass samples. We use numbers 0 to 4 to group the clusters into five approximately distinct categories, across different masses ranges. The cluster C4a/C2 of the most massive sample shows overlapping properties with both clusters C2 and C4 in the fiducial sample, S2, and cluster C4b in the same mass range.}
\end{deluxetable*}

Table~\ref{tbl:agn_frac} quantifies the AGN fractions (divided into Seyferts and LINERs) and star-forming galaxy fractions for the four samples. Throughout this work, we use the [O\,III]/H$\beta$ and [O\,I]/H$\alpha$ line intensity ratios to define the activity types \citep{Kewley+06}. For a cluster of a given sample, we estimate the standard errors of the fractions of the activity types in the cluster as $\sqrt{ f(1-f)/n_c}$ , where $f$ is either the fraction of Seyferts or LINERs or star-forming galaxies in the cluster, and $n_c$ is the sample size of the cluster. For all samples, the C2 and C1 clusters are dominated by star-forming galaxies, C3 by Seyferts, and C4 by LINERs. In other words, there is a broad correspondence between our classification and those of previous work. However, there are significant differences because our clusters contain mixtures of activity types of varying proportions. The aim of the previous classifications was to cleanly separate the activity types based on emission-line ratios using photoionization models or empirical data analysis \citep{Kauffmann+03, Kewley+06,deSouza+17}. In contrast, our aim is to study the correlation among activity type, gas content, SFR, and structural properties. The clear correspondence between the four galaxy populations identified through the SPCA analysis with distinct spectral classes and luminosity states strongly suggests that some physical parameter controls the relative dominance and evolutionary trajectory of the two principal energy mechanisms (star formation and black hole accretion) during the lifecycle of galaxies. What could that be?

\begin{figure*}
\includegraphics[scale=0.3]{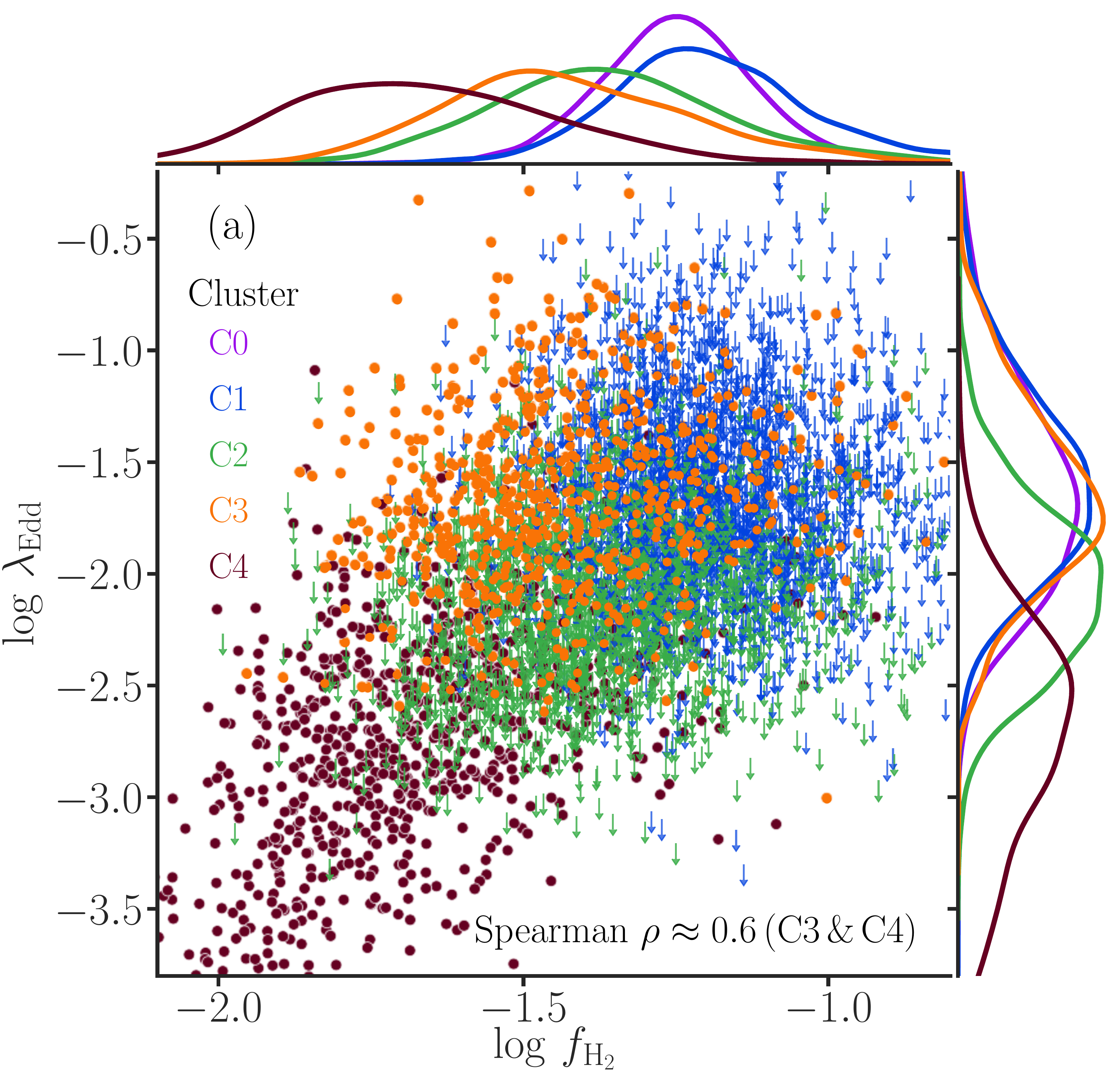}
\includegraphics[scale=0.3]{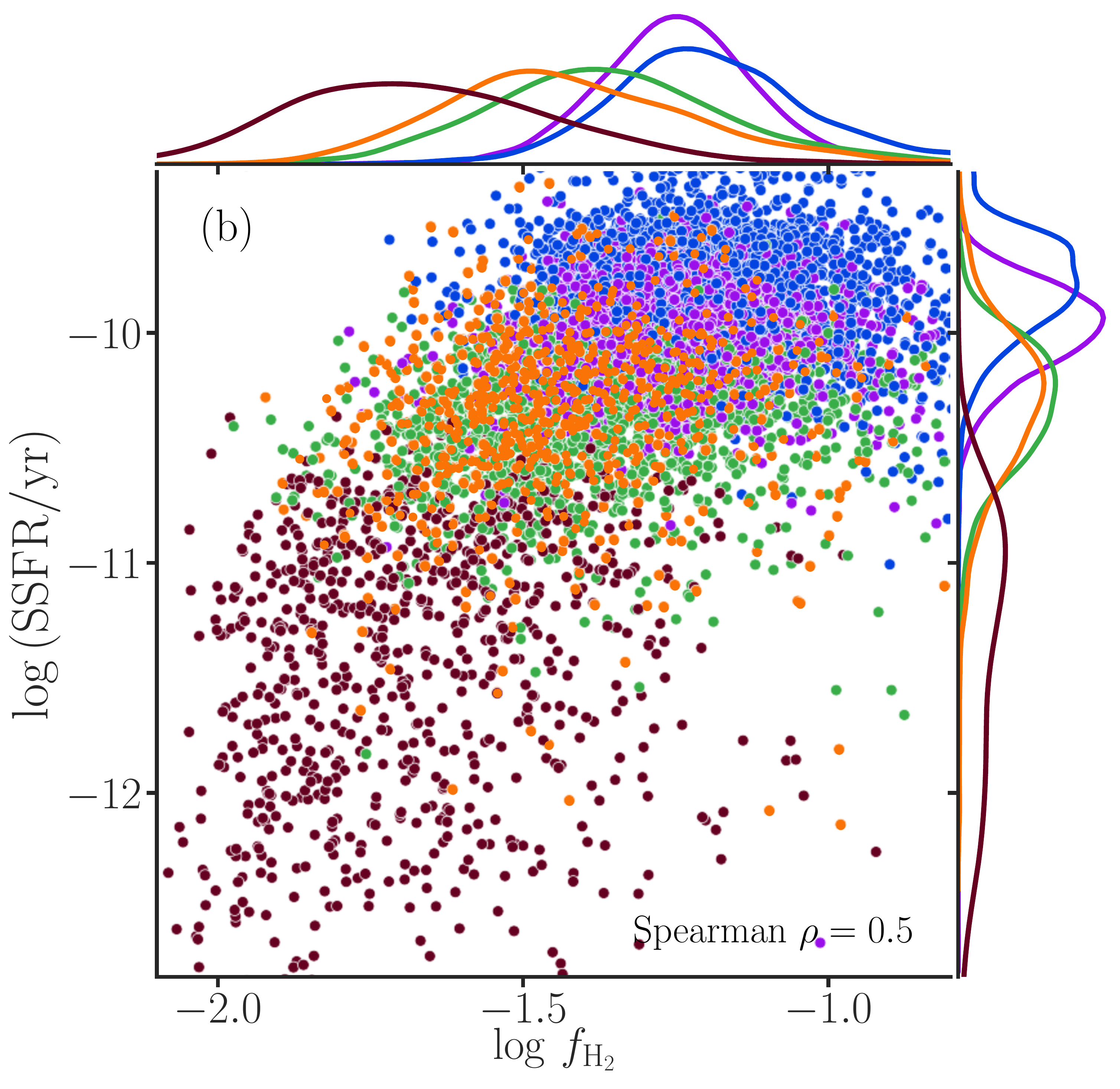}
\includegraphics[scale=0.3]{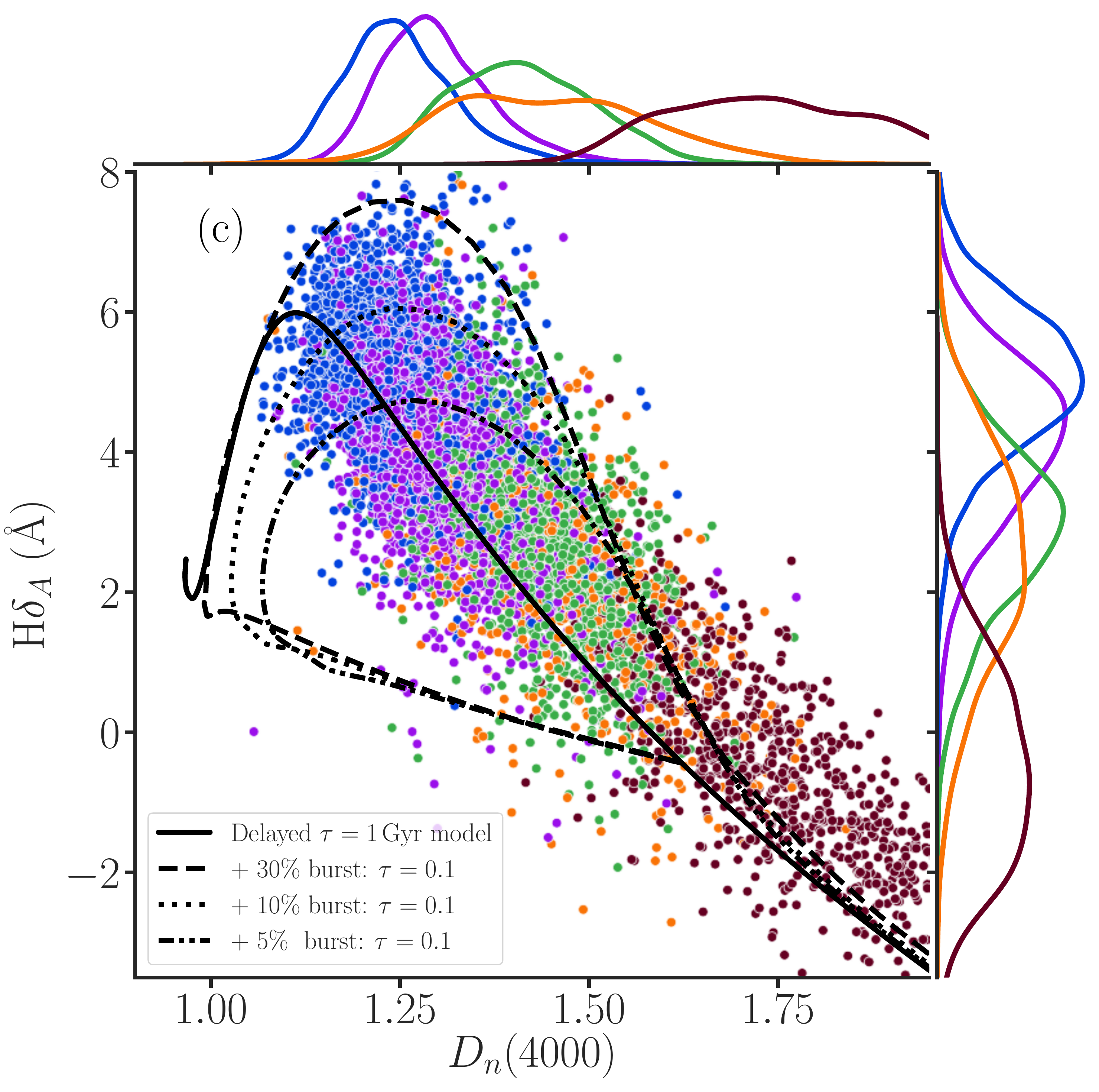}
\caption{Evolution of black hole accretion rate, gas fraction, SFR, and stellar population for galaxies with stellar mass $\log\,(M_\star/M_\odot) = 10.4-10.7$. As before, the colored points are the clusters from our SPCA analysis, and the margin curves depict the kernel density estimates of the quantities plotted. Panel (a) plots the molecular gas fraction ($f_\mathrm{H_2}$) versus the Eddington ratio ($\lambda_\mathrm{Edd}$), revealing a correlation between these two quantities for galaxies in C3 and C4.  Since galaxies in C1 and C2 are dominated by star formation, their $\lambda_\mathrm{Edd}$ are upper limits. The upper limits for C0 are not shown. Panel (b) shows the correlation between $f_\mathrm{H_2}$ and SSFR. Panel (c) plots $D_n(4000)$ and the H$\delta_A$ absorption index, with stellar population models (Bruzual \& Charlot 2003; see Section~2.6) overlaid to indicate how recent bursts of star formation evolve and reproduce the scatter in the diagram. \label{fig:gas_star_corr}} 
\end{figure*}

Answer: gas content.  Figure~\ref{fig:gas_star_corr}a shows that AGN-dominated galaxies (C3 and C4) exhibit a moderately strong relationship (Spearman correlation coefficient $\rho \approx 0.5$) between the molecular gas fraction ($f_\mathrm{H_2} \equiv M_\mathrm{H_2}/M_\star$) and the Eddington ratio, such that more gas results in higher levels of black hole accretion. Seyferts (C3) possess roughly twice as much molecular gas as LINERs (C4). Thus, while the quiescence of LINERs has been attributed to their radiatively inefficient central engines \citep{Ho2008, Ho2009}, their host galaxies also suffer from a starvation diet.  The sizable scatter in Figure~\ref{fig:gas_star_corr}a cautions that the causal link between global gas supply and black hole feeding is not one-to-one, but likely stochastic \citep{Hopkins+06b, Novak+11,Yuan+18}. After all, galaxies in C2 and C3 enjoy similar gas reservoirs ($f_\mathrm{H_2} \approx 4$\%) and yet those in C3 have higher $\lambda_\mathrm{Edd}$; likewise, C1 is more gas-rich than C3 despite having lower $\lambda_\mathrm{Edd}$ on average.  

As expected \citep{Genzel+15,Saintonge+17}, the specific SFR (SSFR $\equiv$ SFR/$M_\star$) of star-forming galaxies scales with $f_\mathrm{H_2}$.  But so, too, do the hosts of bright AGNs (C3), which completely overlap with galaxies with moderate levels of star formation (C2) that lie at the lower end of the blue cloud or main sequence (Figure~\ref{fig:gas_star_corr}b). Indeed, the trend continues to hold even for gas-poor passive galaxies (C4). The observation that AGNs preferentially sit below the main sequence has been routinely (mis)interpreted as evidence for AGN feedback \citep[e.g.,][]{Schawinski+07,Leslie+16}. That C2 and C3 have similar $f_\mathrm{H_2}$ and SSFR, despite their difference in nuclear activity, argues against \emph{instantaneous} negative AGN feedback, perhaps contravening the idea that AGN-driven outflows remove large amounts of gas from galaxies in a short period of time. Galaxies in C1 to C3 likely evolve to C4 when star formation depletes their gas on a timescale $t_\mathrm{dep}  \equiv M_\mathrm{H_2}/\mathrm{SFR} \lesssim 1$ Gyr. Galaxies in C3 and C2 have median (16\%, 84\%) depletions times $\log\, t_\mathrm{dep}/\mathrm{yr} = 8.9\,(8.5, 9.2)$ and $8.9\,(8.7, 9.3)$, respectively. The amount of gas needed to power the Seyfert activity observed in C3 is small ($< 0.03\,M_\odot$~yr$^{-1}$), and even continuous accretion onto the black hole would not deplete $\sim 10^9\,M_\odot$ gas in several Gyr. AGN-driven outflows with mass outflow rates of several $M_\odot$~yr$^{-1}$ can do the job, but the mass outflow rate is highly uncertain \citep{Harrison+18} and difficult to estimate. \citet{Fluetsch+19} found depletion times of molecular gas between few times $10^{6}$ and $10^{8}$\,yr for small samples of local AGNs.

Black hole accretion and star formation are closely intertwined but not coeval. The significant scatter in the $D_n(4000)$--H$\delta_A$ diagram (Figure~\ref{fig:gas_star_corr}c) requires a bursty SFH \citep{Kauffmann+03}. Our toy stellar population models can reproduce the data with recent bursts with mass fractions of $\sim 5\%-30\%$ superposed on an underlying old population. The AGN activity may be linked but delayed with respect to the most recent burst \citep{Kauffmann+03,Yesuf+14} by $\sim 50-250$\,Myr \citep{Davies+07,Wild+10,Yesuf+14}. This is corroborated by the clear separation of the distributions of $D_n(4000)$ and H$\delta_A$ for C1 and C3.  The time delay can be attributed to stellar feedback or dynamical and viscous lags experienced by the gas as it journeys from kpc scales en route to the central black hole \citep{Wild+10,Hopkins+12,Blank+16}.

Tables~\ref{tbl:gas} provides summary statistics for the distributions of H$\alpha$/H$\beta$, gas mass, gas fraction, SFR, the 4000 {\AA} break, morphological parameters, and the numbers of galaxies in the five clusters for each of the four samples. As expected, for a given cluster, SFRs and molecular gas masses increase and gas fractions decrease as the mass range of the sample increases. C2 and C3 have similar gas, stellar and structural properties in all four samples. Their gas mass is $M_\mathrm{H_2} \approx (1-2) \times 10^9\, M_\odot$. This indicates that AGN feedback does not significantly impact cold gas content of galaxies instantaneously. Substantial amounts of cold gas still remain in strong AGNs. Note that the total gas mass, including H\,I, is likely $\sim (4-5) \times M_\mathrm{H_2}$ \citep{Catinella+18}.

\movetabledown=6cm
\begin{rotatetable*}
\begin{deluxetable*}{lcccccccccccc}
\tabletypesize{\scriptsize}
\tablecolumns{13} 
\tablewidth{0pt}
\tablecaption{Gas, Stellar, and Structural Properties of the Clusters in the Different Samples \label{tbl:gas}}
\tablehead{
\colhead{Sample} & \colhead{Cluster} & H$\alpha$/H$\beta$ & \colhead{$\log \,M_\mathrm{H_2}$} & \colhead{$\log \,f_\mathrm{H_2}$} & \colhead{$\log\,\mathrm{SFR}$} & \colhead{$D_n(4000)$} & \colhead{H$\delta$} & \colhead{$C$} & \colhead{$\sigma$} & \colhead{S\'{e}rsic $n$} & \colhead{$\log\,\Sigma_\star$} & \colhead{$N$ Gal.} \\
\colhead{(1)} &
\colhead{(2)} &
\colhead{(3)} &
\colhead{(4)} &
\colhead{(5)} &
\colhead{(6)} &
\colhead{(7)} &
\colhead{(8)} &
\colhead{(9)} &
\colhead{(10)} &
\colhead{(11)} &
\colhead{(12)} &
\colhead{(13)}
}
\startdata
 & C0a &$3.9\,(3.6, 4.3)$ & $9.1\,(8.9, 9.2)$ & $-1.2\,(-1.3, -1.1)$ & $0.2\,(0.1, 0.4)$ & $1.3\,(1.2, 1.4)$ & $4.0\,(2.8, 5.1)$ & $2.1\,(1.9, 2.3)$ & $66\,(49, 82)$ & $0.9\,(0.7, 1.4)$ & $8.3\,(8.1, 8.6)$ & 1478\\ 
 & C0b &$4.1\,(3.8, 4.6)$ & $9.1\,(9.0, 9.2)$ & $-1.1\,(-1.3, -1.0)$ & $0.5\,(0.3, 0.6)$ & $1.24\,(1.2, 1.3)$ & $5.2\,(4.3, 6.0)$ & $2.2\,(2.0, 2.4)$ & $76\,(59, 94)$ & $1.0\,(0.7, 1.5)$ & $8.5\,(8.3, 8.7)$ & 1729 \\ 
S1 & C1 &$4.5\,(3.9, 5.2)$ & $9.1\,(8.9, 9.3)$ & $-1.1\,(-1.3, -1.0)$ & $0.6\,(0.4, 0.8)$ & $1.22\,(1.2, 1.3)$ & $5.4\,(4.3, 6.3)$ & $2.6\,(2.3, 2.8)$ & $92\,(72, 113)$ & $2.1\,(1.2, 4.3)$ & $8.8\,(8.6, 9.0)$ & 938\\ 
 & C2 & $4.1\,(3.6, 4.8)$ & $9.0\,(8.8, 9.2)$ & $-1.3\,(-1.4, -1.1)$ & $0.1\,(-0.1, 0.3)$ & $1.36\,(1.3, 1.4)$ & $3.6\,(2.5, 4.7)$ & $2.6\,(2.4, 2.9)$ & $86\,(68, 106)$ & $2.4\,(1.5, 4.4)$ & $8.8\,(8.6, 9.0)$ & 1139\\ 
 & C3 & $4.0\,(3.6, 4.6)$ & $8.9\,(8.7, 9.1)$ & $-1.3\,(-1.5, -1.2)$ & $0.0\,(-0.3, 0.3)$ & $1.4\,(1.3, 1.6)$ & $2.7\,(0.9, 4.0)$ & $2.6\,(2.2, 2.9)$ & $90\,(69, 112)$ & $2.8\,(1.4, 5.3)$ & $8.8\,(8.6, 9.0)$ & 301\\ 
 & C4 & $3.6\,(3.1, 4.1)$ & $8.7\,(8.5, 8.9)$ & $-1.5\,(-1.7, -1.3)$ & $-0.7\,(-1.5, -0.3)$ & $1.6\,(1.5, 1.8)$ & $0.6\,(-1.1, 2.1)$ & $2.7\,(2.5, 3.1)$ & $95\,(75, 121)$ & $3.6\,(2.2, 5.9)$ & $9.0\,(8.7, 9.2)$ & 385\\   
 \hline
  & C0 & $4.3\,(3.9, 4.7)$ & $9.3\,(9.1, 9.4)$ & $-1.2\,(-1.4, -1.1)$ & $0.5\,(0.3, 0.7)$ & $1.3\,(1.2, 1.4)$ & $4.2\,(2.9, 5.3)$ & $2.1\,(2.0, 2.3)$ & $82\,(65, 101)$ & $1.0\,(0.7, 1.5)$ & $8.5\,(8.3, 8.7)$ & 2824\\ 
  & C1 &$4.7\,(4.2, 5.6)$ & $9.4\,(9.2, 9.5)$ & $-1.2\,(-1.3, -1.0)$ & $0.8\,(0.5, 1.0)$ & $1.24\,(1.2, 1.3)$ & $5.1\,(4.1, 6.0)$ & $2.4\,(2.2, 2.8)$ & $109\,(86, 133)$ & $1.6\,(1.0, 3.1)$ & $8.8\,(8.6, 9.0)$ & 2374\\ 
S2 & C2 &$4.3\,(3.7, 5.0)$ & $9.2\,(9.0, 9.4)$ & $-1.4\,(-1.5, -1.2)$ & $0.2\,(0.0, 0.5)$ & $1.4\,(1.3, 1.5)$ & $3.0\,(1.8, 4.2)$ & $2.7\,(2.4, 3.0)$ & $105\,(86, 128)$ & $2.7\,(1.7, 4.9)$ & $8.9\,(8.7, 9.1)$ & 1808\\ 
 & C3 & $4.3\,(3.8, 5.1)$ & $9.1\,(8.9, 9.3)$ & $-1.4\,(-1.6, -1.2)$ & $0.3\,(-0.1, 0.6)$ & $1.4\,(1.3, 1.6)$ & $2.5\,(0.9, 4.0)$ & $2.7\,(2.3, 3.0)$ & $114\,(94, 138)$ & $3.4\,(1.6, 6.3)$ & $8.9\,(8.7, 9.1)$ & 699 \\ 
 & C4 & $3.6\,(2.9, 4.4)$ & $8.9\,(8.7, 9.1)$ & $-1.7\,(-1.9, -1.5)$ & $-0.7\,(-1.6, -0.2)$ & $1.7\,(1.6, 1.9)$ & $-0.5\,(-2.2, 1.0)$ & $2.9\,(2.6, 3.2)$ & $124\,(101,158)$ & $4.4\,(2.8, 6.5)$ & $9.0\,(8.8, 9.2)$ & 796\\ 
\hline
& C0 &$4.5\,(4.0, 5.0)$ & $9.4\,(9.3, 9.6)$ & $-1.4\,(-1.5, -1.2)$ & $0.8\,(0.5, 1.0)$ & $1.3\,(1.2, 1.4)$ & $3.8\,(2.3, 5.0)$ & $2.2\,(2.0, 2.4)$ & $104\,(82, 127)$ & $1.2\,(0.8, 2.0)$ & $8.6\,(8.5, 8.8)$ & 2509\\ 
 & C1 & $5.3\,(4.6, 6.3)$ & $9.6\,(9.4, 9.8)$ & $-1.3\,(-1.5, -1.1)$ & $1.0\,(0.7, 1.2)$ & $1.3\,(1.2, 1.4)$ & $4.8\,(3.6, 5.9)$ & $2.6\,(2.3, 2.9)$ & $139\,(114, 166)$ & $2.1\,(1.3, 4.1)$ & $8.9\,(8.7, 9.1)$ & 1991\\ 
 S3 & C2 & $4.5\,(3.9, 5.2)$ & $9.3\,(9.2, 9.5)$ & $-1.5\,(-1.7, -1.3)$ & $0.4\,(0.1, 0.7)$ & $1.5\,(1.4, 1.6)$ & $2.5\,(1.0, 3.8)$ & $2.8\,(2.5, 3.1)$ & $131\,(108, 156)$ & $3.6\,(2.2, 6.1)$ & $8.9\,(8.7, 9.1)$ & 1969\\ 
 & C3 &$4.3\,(3.8, 5.0)$ & $9.3\,(9.1, 9.5)$ & $-1.6\,(-1.8, -1.4)$ & $0.4\,(0.1, 0.7)$ & $1.5\,(1.3, 1.6)$ & $2.3\,(0.3, 3.9)$ & $2.7\,(2.3, 3.0)$ & $134\,(107, 160)$ & $3.4\,(1.8, 6.2)$ & $8.9\,(8.7, 9.1)$ & 1061\\ 
& C4 &$3.6\,(2.9, 4.3)$ & $9.0\,(8.8, 9.2)$ & $-1.9\,(-2.1, -1.6)$ & $-0.6\,(-1.4, 0.0)$ & $1.8\,(1.7, 1.9)$ & $-1.1\,(-2.3, 0.3)$ & $3.0\,(2.6, 3.3)$ & $152\,(121, 188)$ & $4.6\,(2.9, 6.5)$ & $9.0\,(8.8, 9.2)$ & 1345\\ 
\hline
& C0 & $4.5\,(3.9, 5.1)$ & $9.5\,(9.3, 9.7)$ & $-1.6\,(-1.8, -1.4)$ & $0.7\,(0.4, 1.0)$ & $1.4\,(1.3, 1.6)$ & $2.4\,(0.5, 3.8)$ & $2.4\,(2.1, 2.6)$ & $127\,(106, 154)$ & $2.1\,(1.2, 3.1)$ & $8.8\,(8.6, 9.0)$ & 861\\ 
 & C1 & $5.5\,(4.7, 7.0)$ & $9.7\,(9.5, 10)$ & $-1.4\,(-1.6, -1.1)$ & $0.9\,(0.6, 1.2)$ & $1.35\,(1.3, 1.4)$ & $4.0\,(2.8, 5.3)$ & $2.7\,(2.4, 3.0)$ & $164\,(136, 191)$ & $2.8\,(1.7, 4.7)$ & $9.0\,(8.9, 9.2)$ & 758\\ 
S4 & C3 &$4.3\,(3.8, 5.0)$ & $9.3\,(9.2, 9.6)$ & $-1.8\,(-1.9, -1.6)$ & $0.5\,(0.1, 0.8)$ & $1.6\,(1.4, 1.7)$ & $1.0\,(-0.8, 2.7)$ & $2.7\,(2.3, 3.0)$ & $153\,(127, 179)$ & $3.5\,(2.2, 5.9)$ & $8.9\,(8.8, 9.1)$ & 487\\ 
 & C4a/C2 & $4.5\,(3.8, 5.4)$ & $9.4\,(9.2, 9.6)$ & $-1.7\,(-2.0, -1.5)$ & $0.2\,(-0.5, 0.6)$ & $1.6\,(1.5, 1.8)$ & $0.6\,(-1.2, 2.4)$ & $3.0\,(2.7, 3.2)$ & $175\,(147, 202)$ & $4.4\,(3.0, 6.5)$ & $9.0\,(8.9, 9.2)$ & 803\\ 
 & C4b & $3.4\,(2.7, 4.1)$ & $9.0\,(8.9, 9.3)$ & $-2.1\,(-2.3, -1.9)$ & $-0.7\,(-1.4, 0.0)$ & $1.9\,(1.8, 2.0)$ & $-1.6\,(-2.5, -0.6)$ & $3.1\,(2.7, 3.3)$ & $188\,(153, 225)$ & $4.7\,(3.3, 6.6)$ & $9.0\,(8.8, 9.2)$ & 883\\ 
\enddata 
\tablecomments{
(1) Narrow mass samples. (2) Cluster names from the SPCA+clustering analysis. (3) Balmer decrement. (4) Molecular gas mass in $M_\odot$ (5) Gas fraction, $f_\mathrm{H_2} =  \,M_\mathrm{H_2}/M_\star$. (6) Star formation rate in $M_\odot\,\mathrm{yr}^{-1}$. (7) The 4000\, {\AA} break. (8) The equivalent width of H$\delta$ absorption in {\AA}. (9) Concentration index, $C = R_{90}/R_{50}$, (10) Stellar velocity dispersion in km\,s$^{-1}$. (11) Global S\'{e}rsic index. (12) Stellar mass density in $M_\odot\,\mathrm{kpc}^{-2}$ (13) Number of galaxies in a cluster. We use the notation $X\,(Y, Z)$ to denote $X =$ median (50\%), $Y = 16\%$, and $Z= 84\%$ of a distribution.}
\end{deluxetable*}
\end{rotatetable*}

Table~\ref{tbl:rho_corr} presents the Spearman correlation coefficients ($\rho$) for the quantities plotted in Figure~\ref{fig:gas_star_corr} and for the four samples. We provide $\rho$ values of the relations including all four clusters, only the AGN-dominated clusters C3 and C4, and only using AGNs classification according to Figure~\ref{fig:bpt_evol}. In all four samples, AGNs show moderate ($\rho \approx 0.5-0.7$) correlations between $f_\mathrm{H_2}$ and SSFR, and between $f_\mathrm{H_2}$ and $\lambda_\mathrm{Edd}$. 

\section{DISCUSSION}\label{sec:disc}

\subsection{Strong AGNs are gas-rich and dusty}

In \citet{Yesuf+19}, we showed that dust absorption is a dirt-cheap method to estimate molecular gas masses for large samples of galaxies. This method can be applied to a variety of problems. Here, we use it to study $\sim 27,100$ central, massive galaxies in narrow stellar mass ranges in order to identify galaxy populations that are likely to be evolutionarily related. We identify four galaxy groups that can be evolutionarily linked through a lifecycle wherein gas content modulates their SFR and level of AGN activity. We find that strongly accreting black holes (Seyferts) live in gas-rich, young, and star-forming hosts, unlike their weakly accreting counterparts (LINERs). Thus, we do not find evidence for \emph{rapid or instantaneous} impact of AGN feedback on the molecular gas and dust in nearby galaxies. Our result supports the observational  \citep{Netzer09,Diamond-Stanic+12,Esquej+14,Delvecchio+15,Izumi+16,Shimizu+17,Shangguan20,ZhuangHo20} and theoretical \citep{Hopkins+10,Thacker+14, Volonteri+15,McAlpine+17} claim that strong AGNs co-habitate in (gas-rich) star-forming galaxies.

In a companion paper (Yesuf \& Ho 2020, submitted), we also apply our method to study a large sample of post-starburst galaxies (PSBs). They are rare but important class of galaxies that are rapidly quenching. We find that large reservoirs of molecular gas are present in significant numbers of post-starburst AGNs, and that they are not removed or destroyed by AGN feedback in some PSBs. Next, we discuss the nature of outflows in the AGNs in the current sample, and whether AGNs are preferentially powered by bar-driven inflows.

\subsection{AGN-driven outflows}

Simulations have shown that AGN feedback may drive strong outflows with little impact on the dense gas in the galaxy disk \citep{Gabor+14, Roos+15}. AGNs mainly affect the diffuse gas in the interstellar and CGM. AGN-driven outflows in these simulations do not cause rapid quenching of star formation, but they may remove significant amounts of gas over long timescales ($> 10^9$ yr). It is widely accepted that AGN feedback plays an important role in keeping the CGM hot and in maintaining long-term quenching \citep{Croton+06,Fabian+12}. Although based on small samples and uncertain assumptions, \citet{Fluetsch+19} found a scaling relation with the bolometric luminosity of AGN or SFR that predict average gas outflow rates in AGN hosts. Using these relations with median $\log\,\mathrm{SFR}/M_\odot$~yr$^{-1}$ = 0.3 or median $\log\,L_\mathrm{bol}/\mathrm{erg\,s^{-1}} = 43.8$, we estimate an outflow rate for the C3 cluster in the S2 sample of $\sim 11\,M_\odot$~yr$^{-1}$ or $\sim 20\,M_\odot$~yr$^{-1}$, respectively. These estimates seem too high for the observed change in gas mass from C3 to C4 of $\sim 5 \times 10^8\, M_\odot$. If the average time between C3 and C4 is $ \Delta t \gtrsim 10^8$ years, then the AGN outflow rate is $\lesssim 5\,M_\odot$ \,yr$^{-1}$, even ignoring the gas consumption by a SFR of $\sim 2\,M_\odot$~yr$^{-1}$. Modeling the observed SFR by a SFH $\psi  \propto \exp(-t/\tau)$ with $\tau = 100 - 200$\,Myr, $\Delta t = 1.8\,\tau$. Likewise, in rapidly quenching galaxies, the time interval between the Seyfert and LINER post-starburst phases is $\gtrsim 3 \times 10^8$ yr \citep{Yesuf+14}. Furthermore, the AGN lifetime is likely $\sim 5 \times 10^7- 5 \times 10^8$~yrs for AGNs similar to those in C3 \citep{Martini+01,Goncalves+08, Hopkins+09, Borisova+16}. Thus, while the current work cannot rule out significant gas loss due to AGN-driven outflows, we do not find a compelling reason to invoke it, in addition to gas consumption by star formation. Likewise, \citet{Jarvis+20} recently showed that some $z \sim 0.1$ type 2 quasars with strong, kpc-scale ionized gas outflows and jets are star-forming and gas rich, and that there is no evidence for immediate appreciable impact of AGN feedback on the global molecular gas reservoirs of these powerful AGNs. Next, we briefly discuss the role of AGNs in regulating new gas accretion according to cosmological simulations \citep{Davies+20, Oppenheimer+20,Terrazas+20, Zinger+20}.

\begin{deluxetable}{lcc}
\tabletypesize{\footnotesize}
\tablecolumns{3} 
\tablewidth{0pt}
\tablecaption{Spearman Correlation Coefficients \label{tbl:rho_corr}}
\tablehead{
\colhead{Sample} & \colhead{$f_\mathrm{H_2}$ vs. SSFR} & \colhead{$f_\mathrm{H_2}$ vs. $\lambda_\mathrm{Edd}$}
}
\startdata
S1 without C0 & 0.45 & $\sim 0.3$ \\
S2 without C0 & 0.54 & $\sim 0.4$\\
S3 without C0 & 0.63 & $\sim 0.5$\\
S4 without C0 & 0.66 & $\sim 0.7$ \\
\hline 
S1 C3 and C4 only & 0.44 & 0.57\\
S2 C3 and C4 only & 0.51 & 0.61 \\
S3 C3 and C4 only & 0.54 & 0.66\\
S4 C3 and C4b only & 0.59 & 0.73\\
S4 C3, C4a and C4b only & 0.51 & 0.66\\
\hline
S1 AGNs only & 0.44 & 0.62 \\
S2 AGNs only & 0.47 & 0.60\\
S3 AGNs only & 0.48 & 0.63 \\
S4 AGNs only & 0.49 & 0.70\\
\enddata
\end{deluxetable}

\begin{figure*}
\includegraphics[scale=0.45]{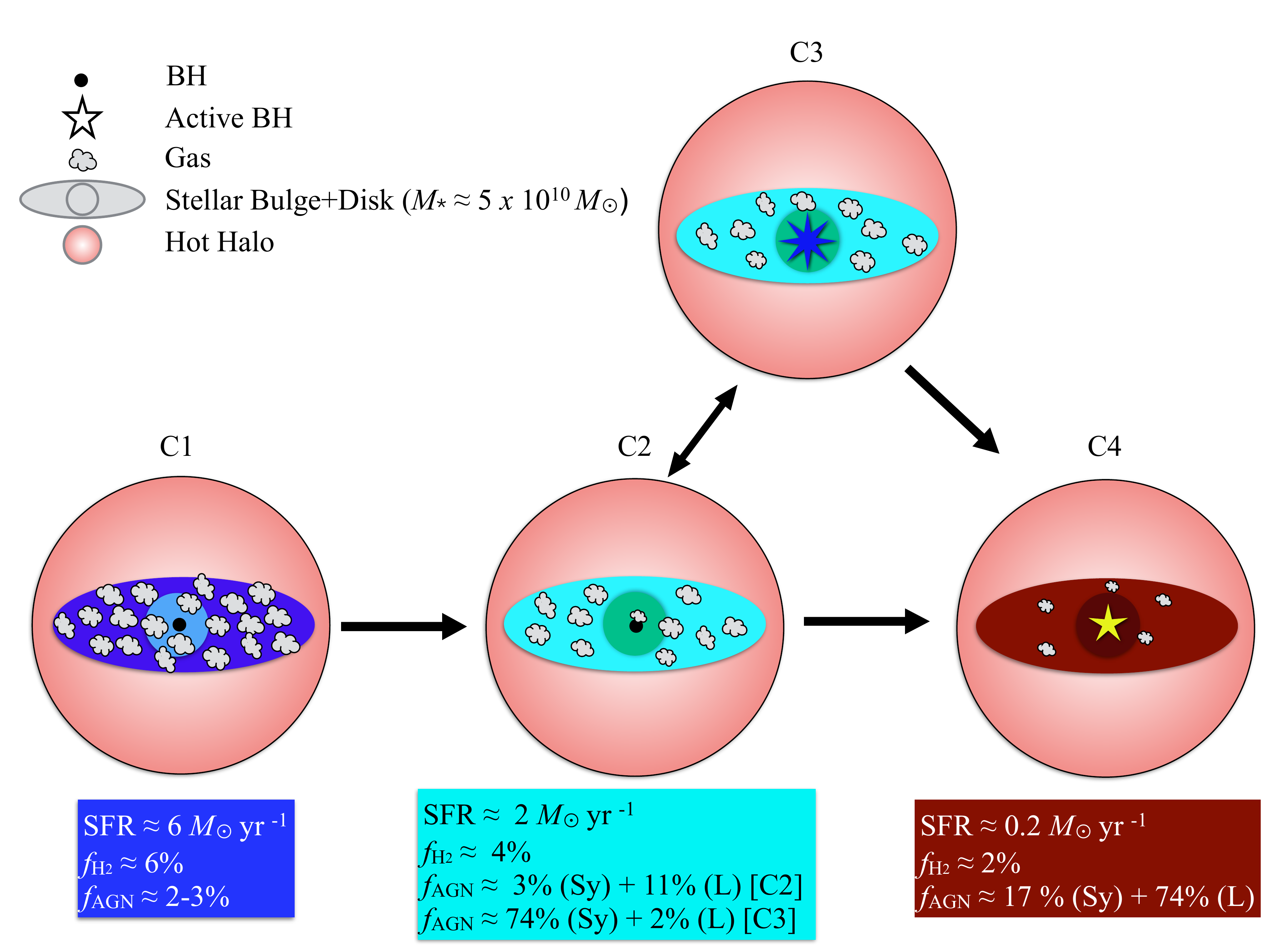}
\caption{Schematic diagram summarizing the results of this study. We identify four ``clusters'' (C1 to C4) of central, massive, and bulge-dominated galaxies that are evolutionarily linked. The lifecycle begins with C1 and ends with C4, with the SFR and molecular gas fraction decreasing along the sequence. The intermediate-stage clusters C2 and C3 have similar gas and stellar properties, but C3 is dominated by strong AGNs while C2 has much fewer AGNs. The double-headed arrow connecting C2 and C3 indicates the possibility that galaxies may go back and forth between these two groups as the AGNs flicker. The most powerful AGNs (Seyferts) emerge during phase C3, terminating as LINERs in C4. The molecular gas is depleted within $\sim 1$ Gyr mainly by star formation, with negligible impact from AGN feedback. \label{fig:cartoon}}
\end{figure*}

Although there are quantitative differences between them, both EAGLE and TNG simulations indicate that the expulsion of rapidly cooling gas may lead to a long CGM cooling time, effectively depleting the halo of gas that would otherwise replenish the interstellar medium and fuel future star formation \citep{Davies+20}. In both sets of simulations, central halos with high CGM gas fractions ($f_\mathrm{CGM}$) preferentially host galaxies that are more star-forming and have greater rotational support than typical halos. In addition, halos with lower $f_\mathrm{CGM}$ at fixed mass have more massive black holes and higher ratios of cumulative black hole feedback energy to halo binding energy \citep{Oppenheimer+20,Terrazas+20}. At a fixed halo mass, $f_\mathrm{CGM}$ is lower in the EAGLE simulations because the black holes, which are hosted by early-assembled  halos, reach high accretion rates sooner. In contrast, in TNG simulations $f_\mathrm{CGM}$ is lower at fixed mass when the central black holes reach the mass threshold at which AGN feedback switches from the thermal to kinetic mode. In EAGLE simulations, the scatter about the median $f_\mathrm{CGM}$ is not correlated with the AGN luminosity/accretion rate, but TNG simulations predict strong correlation between these quantities, especially for the characteristic halo mass $M_{h} \approx 10^{12}\,M_\odot$. How these differences translate to the molecular gas in the interstellar medium and which simulations agree better with the trends presented in this work are yet to be seen in future studies. To interpret our results in the meantime, both simulations indicate that the CGM cooling times for the present-day galaxies near the characteristic mass are $\gtrsim 1$\,Gyr \citep{Davies+20, Zinger+20}. Hence, it may be assumed that the net inflow rates to the interstellar medium are not changing rapidly as local massive galaxies quench. The evolution of gas fraction already in the interstellar medium governs the observed rates of star formation and black hole accretion.
 
\subsection{The effect of galactic bars}

Stellar bars can efficiently transfer angular momentum and instigate radial inflow of gas to trigger nuclear activity \citep{Shlosman+89}. However, the four evolutionary classes defined in our study have similar bar fractions (Table~\ref{tbl:fbar}). Within sample S2, the bar fractions are $f_\mathrm{bar} \approx 0.30 \pm 0.02$ for clusters C1 to C3, $0.25 \pm 0.01$ for C4, and $0.38 \pm 0.01$ for C0. Bar classifications using low resolution images are highly uncertain. The values of $f_\mathrm{bar}$ may increase by up to $\sim 10$\% for $z < 0.07$. Similar to our results, previous attempts to connect AGN fueling to bars have yielded negative \citep{Ho+97b,Cisternas+13,Cheung+15,Galloway+15,Goulding+17,Neumann+19} or at best ambiguous \citep{Laine+02,Alonso+18} results.

\subsection{Caveats} Our method can only predict gas masses within a factor of 2.5 \citep{Yesuf+19}. For this reason, the intrinsic correlation of AGN activity or star formation with gas fraction may be stronger than reported here. Furthermore, we can only estimate the global gas content. We do not have information on the interstellar medium on nuclear scales. Lastly, our current study does not include quasars and unobscured, type 1 AGNs. It is unclear whether our main conclusions extend to these AGN populations. We should bear in mind, however, that a recent far-infrared study indicates that the gas content of the host galaxies of type 1 and type 2 quasars are similar, and neither type is gas-deficient relative to normal, inactive galaxies \citep{Shangguan+19}.  This is supported by the work of \citet{ZhuangHo20}, which is also based on the method of \citet{Yesuf+19}.

\begin{deluxetable}{lccc}
\tabletypesize{\footnotesize}
\tablecolumns{4} 
\tablewidth{0pt}
\tablecaption{Bar Fractions \label{tbl:fbar}}
\tablehead{
\colhead{Sample} & \colhead{Clusters } & \colhead{$f_\mathrm{bar}$} & \colhead{$f_\mathrm{bar}$} ($z < 0.07$)}
\startdata
 & C0a & $0.39 \pm 0.01$ & $0.41 \pm 0.02$\\ 
 & C0b & $0.36 \pm 0.01$ & $0.39 \pm 0.02$ \\ 
S1 & C1 &$0.26 \pm 0.01$ & $0.26 \pm 0.02$ \\ 
 & C2 & $0.27 \pm 0.01$ & $0.27 \pm 0.02$ \\ 
 & C3 & $0.29 \pm 0.02$ & $0.31 \pm 0.03$ \\ 
 & C4 & $0.23 \pm 0.02$ & $0.23 \pm 0.02$ \\   
   \hline
  & C0 & $0.38 \pm 0.01$ & $0.47 \pm 0.02$ \\ 
  & C1 & $0.31 \pm 0.01$ & $0.34 \pm 0.02$\\ 
S2 & C2 &$0.31 \pm 0.01$ & $0.34 \pm 0.02$\\ 
 & C3 & $0.30 \pm 0.02$ & $0.35 \pm 0.03$\\ 
 & C4 & $0.25 \pm 0.01$ & $0.25 \pm 0.02$\\ 
\hline
& C0 & $0.40 \pm 0.01$ & $0.55 \pm 0.03$ \\ 
  & C1 & $0.31 \pm 0.01$ & $0.42 \pm 0.04$\\ 
S3 & C2 &$0.34 \pm 0.01$ & $0.41 \pm 0.02$\\ 
 & C3 & $0.36 \pm 0.01$ & $0.41 \pm 0.04$\\ 
 & C4 & $0.27 \pm 0.01$ & $0.29 \pm 0.02$\\ 
 \hline
& C0 &  $0.49 \pm 0.02$  & $0.61 \pm 0.04$\\ 
 & C1 & $0.35 \pm 0.02$ & $0.48 \pm 0.06$\\ 
S4 & C3 & $0.44 \pm 0.02$ & $0.56 \pm 0.06$ \\ 
 & C4a/C2 & $0.29 \pm 0.01$ & $0.30 \pm 0.04$ \\ 
 & C4b & $0.25 \pm 0.01$ & $0.28 \pm 0.02$\\ 
\enddata
\end{deluxetable}

\section{SUMMARY AND CONCLUSIONS}\label{sec:conc}

Utilizing our recent method to estimate the molecular gas mass using dust absorption \citep{Yesuf+19}, we study the connection between gas content, star formation, and black hole accretion in $\sim 27,100$ face-on, central galaxies at $z = 0.02-0.15$ and with stellar mass $M_\star \approx 10^{10}-2\times 10^{11}\,M_\odot$. Principal component and clustering analysis of a large number of physical observables reveal four populations of galaxies that are evolutionarily linked. 

Figure~\ref{fig:cartoon} presents a schematic, unifying picture of how the availability of interstellar gas affects the joint evolution of star formation and black hole accretion. Gas-rich galaxies first consume their fuel through vigorous star formation (C1). This is followed by an intermediate phase during which moderate gas supply sustains a reduced rate of star formation (C2) in concert with strong Seyfert activity mediated by stochastic accretion (C3). AGN feedback, however, does not seem to have instantaneous, observable effects on the molecular gas content or SFR. Finally, gas exhaustion extinguishes star formation and reduces black hole accretion to the levels seen in LINERs (C4).

AGN-dominated galaxies (C3 and C4) exhibit a moderately strong relationship (Spearman $\rho \approx 0.6$) between the molecular gas fraction and the Eddington ratio, such that more gas results in higher levels of black hole accretion. Seyferts live in young, star-forming, and gas-rich ($M_\mathrm{H_2} \approx 10^9\, M_\odot$) hosts, with median (16\%, 84\%) depletion times $\log\, t_\mathrm{dep}/\mathrm{yr} = 8.9\,(8.5, 9.2)$.

Our study adopts a data-driven approach and is not meant to test specific AGN feedback models. Future comparisons with our results may be useful to discriminate between different models. Our results would be inconsistent with AGN feedback models that predict that central, bulge-dominated, Seyfert-like AGNs in massive galaxies have significantly lower molecular gas fractions compared to inactive galaxies of similar mass, morphology, and SFR. In addition, future observational work can expand upon the current work by (1) directly measuring the molecular gas masses in a carefully selected, representative sample of galaxies (including low-luminosity AGNs and low-SFR AGNs) and repeating the correlations reported here, (2) estimating resolved gas masses either directly or indirectly and studying the effects of AGNs on the gas in nuclear regions, (3) using hard X-ray luminosity, which is less affected by strong star formation, instead of {\LOIII} as a proxy for black hole accretion rate, and (4) utilizing bar classifications based on higher resolution images than those provided by SDSS to robustly study the connection between AGN fueling and bars. 

\acknowledgements

We are very thankful to the anonymous referee for the helpful feedback, comments and suggestions that significantly improved the content and presentation of the paper.

This work was supported by the National Science Foundation of China (11721303, 11991052, 11950410492) and the National Key R\&D Program of China \,(2016YFA0400702). 

Funding for SDSS has been provided by the Alfred P. Sloan Foundation, the Participating Institutions, the National Science Foundation, and the U.S. Department of Energy Office of Science. The SDSS-III web site is http://www.sdss3.org/.  
SDSS-III is managed by the Astrophysical Research Consortium for the Participating Institutions of the SDSS-III Collaboration including the University of Arizona, the Brazilian Participation Group, Brookhaven National Laboratory, Carnegie Mellon University, University of Florida, the French Participation Group, the German Participation Group, Harvard University, the Instituto de Astrofisica de Canarias, the Michigan State/Notre Dame/JINA Participation Group, Johns Hopkins University, Lawrence Berkeley National Laboratory, Max Planck Institute for Astrophysics, Max Planck Institute for Extraterrestrial Physics, New Mexico State University, New York University, Ohio State University, Pennsylvania State University, University of Portsmouth, Princeton University, the Spanish Participation Group, University of Tokyo, University of Utah, Vanderbilt University, University of Virginia, University of Washington, and Yale University.


\appendix 

\section{Additional diagnostics}

\begin{figure*}
\includegraphics[scale=0.38]{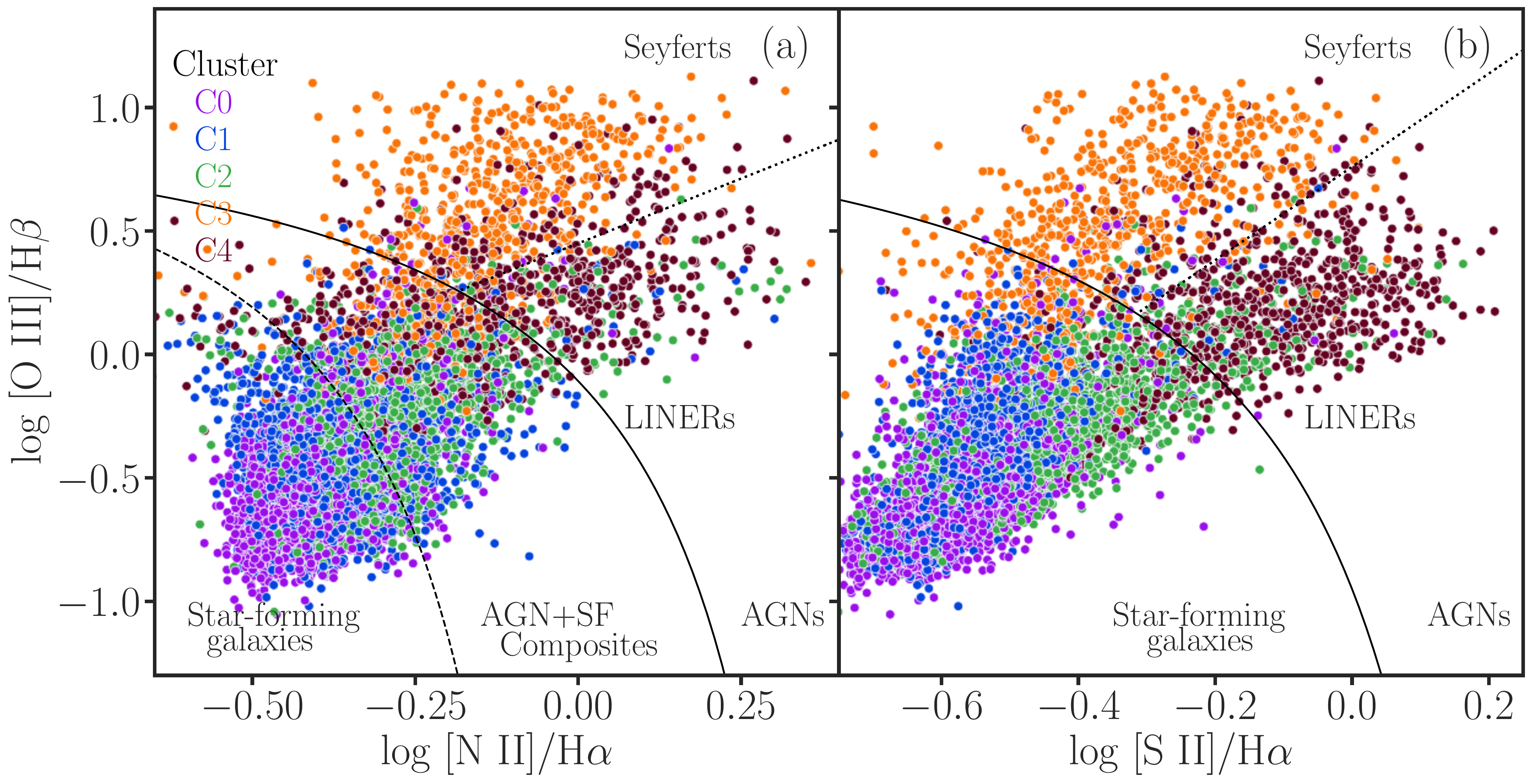}
\caption{Similar to Figure~\ref{fig:bpt_evol}. The solid curve marks the theoretical maximum boundary for star-forming galaxies \citep{Kewley+01}, while the dashed curve represents the empirical boundary of pure star-forming galaxies \citep{Kauffmann+03}; composite galaxies between the two curves have both star formation and AGN activity. The dotted line in the AGN region separates Seyferts from LINERs \citep{Kewley+06, Schawinski+07}. \label{fig:bpt_evol2}}
\end{figure*}

For completeness, Figure~\ref{fig:bpt_evol2} plots the [\ion{N}{2}]~$\lambda 6584$/H$\alpha$ and [\ion{S}{2}]~$\lambda\lambda 6716, 6731$/H$\alpha$ diagrams \citep{Kewley+01,Kauffmann+03,Kewley+06, Schawinski+07} color-coded by the four clusters. The results are similar to Figure~\ref{fig:bpt_evol}a. AGN-dominated galaxies occupy the upper-right corners of the diagrams, while star-forming galaxies occupy the lower-left corners. In the [\ion{N}{2}]/H$\alpha$ diagram, composite/transition galaxies with some mixture of AGN and star formation activity lie in between \citep{Ho+93}. The composite region is populated by all four classes. In other words, composite galaxies are heterogeneous and span a wide range in star formation, age, and gas content.

\begin{figure*}
\includegraphics[scale=0.28]{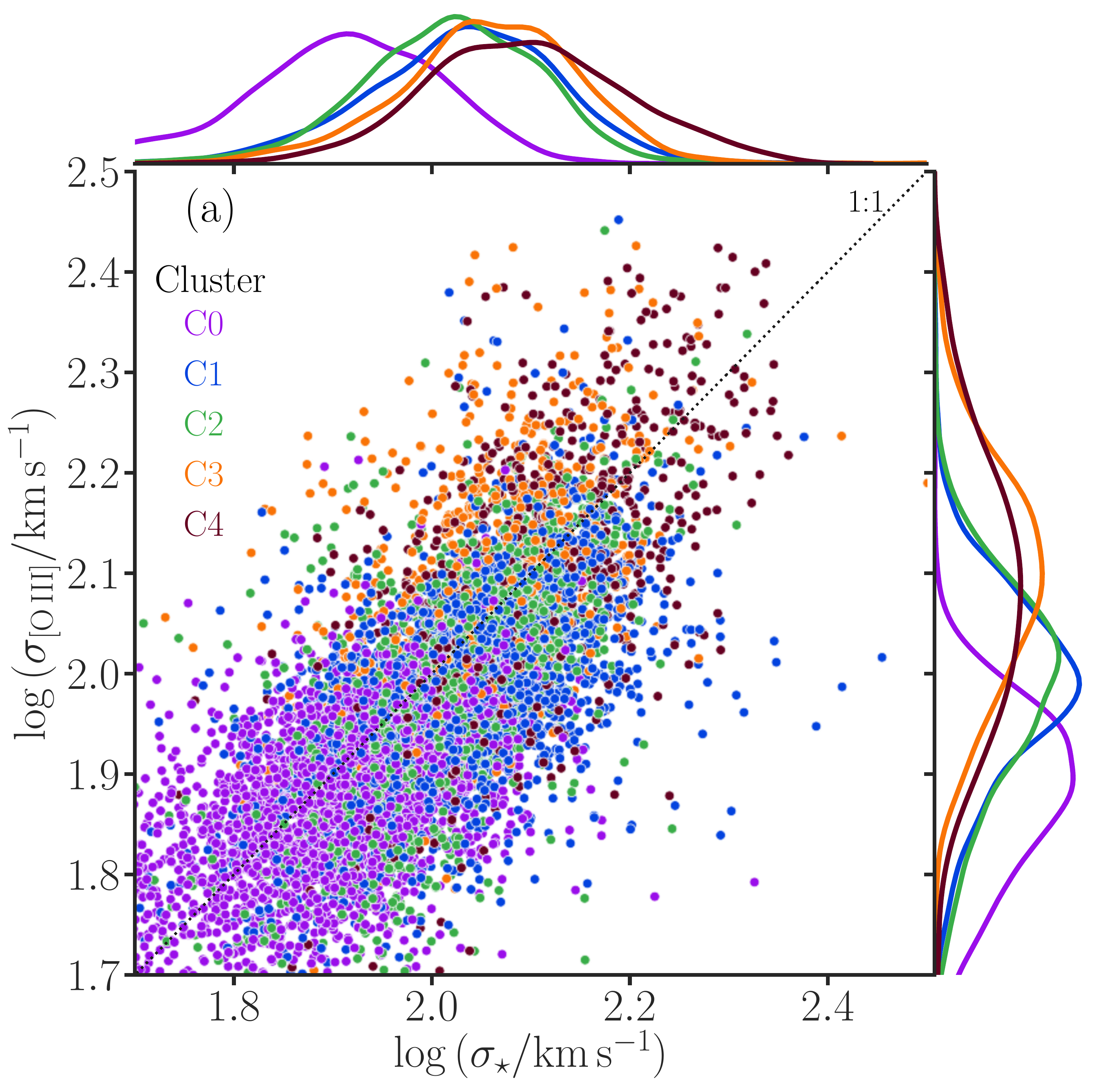}
\hfill
\includegraphics[scale=0.28]{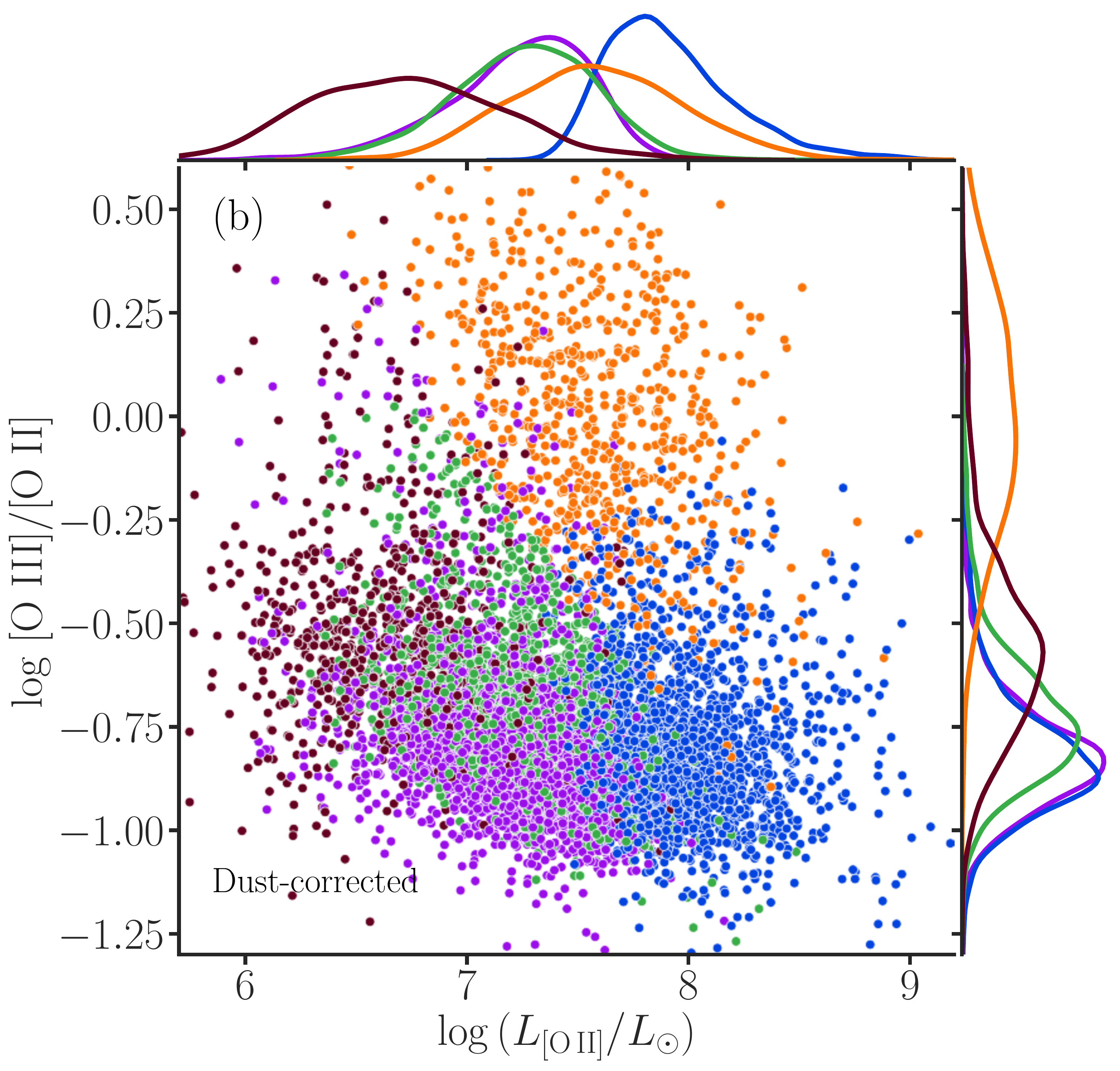}
\caption{Evolution of ionized gas velocity and ionization ratio. Panel (a) compares the velocity dispersions of stars ($\sigma_\star$) and ionized gas ($\sigma_{\rm [O~III]}$). In panel (b), C3 is distinguishable from C1 and C2 because it has stronger ionization parameter ([O\,III]/[O\,II]) from black hole accretion.  The SFRs ({\LOII}) of C1 are higher than those of C2 and C0.\label{fig:extra}}
\end{figure*}

Figure~\ref{fig:extra}a shows that, overall, the stellar velocity dispersion is correlated ($\rho = 0.7$) with gas velocity dispersion measured using all forbidden lines simultaneously. We simply denote it as $\sigma_\mathrm{[O\,III]}$. This correlation is also observed in previous work \citep{GreeneHo05,Ho09b}. Galaxies in C3, however, exhibit more turbulent gas motions relative to their stellar velocity dispersions. This may be due to AGN-driven outflows \citep{Ho09b,KongHo2018}, which are commonly observed in AGN hosts \citep[see][and references therein]{Fiore+17,Yesuf+17b,Harrison+18,Yesuf+20b}. However, without spatially resolved kinematics, the possibility of inflows of low-angular momentum gas powering nuclear activity in C3 galaxies cannot be ruled out \citep{Ho+03}. High spatial resolution data show that both inflow and outflow can co-exist in the same AGN host \citep{Audibert+19}. The fact that the distributions of stellar velocity dispersions for the four clusters are similar implies that their black hole mass distributions are similar \citep{Kormendy+13}. Their inferred median (16\%, 84\%) black hole mass is $\log\,(M_\bullet/M_\odot) \approx 7.4\,(6.9, 7.7)$.

Figure~\ref{fig:extra}b shows dust-corrected [\ion{O}{2}] luminosity and [\ion{O}{3}]/[\ion{O}{2}] ratio. This figure aims to particularly show that the [\ion{O}{2}] and [\ion{O}{3}] luminosities of C1 galaxies are consistent with ionization by newly formed, massive stars, and that the [\ion{O}{2}] luminosities of these galaxies agree with the expectation from their SFRs \citep{ZhuangHo19} given in Table~\ref{tbl:gas}.

\end{document}